\documentclass[aps,prb,twocolumn,groupedaddress]{revtex4-1}

\usepackage{graphicx}
\usepackage{amssymb}
\usepackage{amsmath}
\usepackage{bm}
\usepackage{hyperref}
\usepackage{lipsum}
\usepackage{bbold}
\usepackage{babel}
\usepackage{capt-of}
\usepackage{multirow}
\usepackage{comment} 
\usepackage{multirow}

\begin{document}

\title{
Electron-hole asymmetry and band gaps of commensurate double moire patterns in twisted bilayer graphene on hexagonal boron nitride
}

\author{Jiseon \surname{Shin}}
\affiliation{Department of Physics, University of Seoul, Seoul 02504, Korea}
\author{Youngju \surname{Park}}
\affiliation{Department of Physics, University of Seoul, Seoul 02504, Korea}
\author{Bheema Lingam \surname{Chittari}}
\affiliation{Department of Physics, University of Seoul, Seoul 02504, Korea}
\author{Jeil \surname{Jung}}
\email[jeiljung@uos.ac.kr]{}
\affiliation{Department of Physics, University of Seoul, Seoul 02504, Korea}
\affiliation{Department of Smart Cities, University of Seoul, Seoul 02504, Korea}

%\date[]{Received 13 July 2020}

\begin{abstract}
Spontaneous orbital magnetism observed in twisted bilayer graphene (tBG) on nearly aligned hexagonal boron 
nitride (BN) substrate builds on top of the electronic structure resulting from combined G/G and G/BN double moire interfaces. 
Here we show that tBG/BN commensurate double moire patterns can be classified into two types,
each favoring the narrowing of either the conduction or valence bands on average, 
and obtain the evolution of the bands as a function of the interlayer sliding vectors and electric fields.
Finite valley Chern numbers $\pm 1$ are found in a wide range of parameter space when the moire bands are isolated through gaps,
while the local density of states associated to the flat bands are weakly affected by the BN substrate invariably concentrating around the AA-stacked regions of tBG.
We illustrate the impact of the BN substrate for a particularly pronounced electron-hole asymmetric 
band structure by calculating the optical conductivities of twisted bilayer graphene near the magic angle as a 
function of carrier density.
The band structures corresponding to other $N$-multiple commensurate moire period ratios indicate 
it is possible to achieve narrow width $W \lesssim 30$~meV isolated folded band bundles for tBG angles $\theta \lesssim 1^{\circ}$.
\end{abstract}

%\pacs{68.37.Ef, 82.20.-w, 68.43.-h}

\maketitle

%------------------------------------------------------------%
\section{Introduction}\label{intro}
The electronic properties of magic angle twisted bilayer graphene (tBG) and other graphene moire systems
have become an intense focus of research in recent years after observations of strong 
correlations and superconductivity in transport experiments~\cite{mott1,mott2,mott3, yankowitz2019tuning, macdonald2019trend, andreayoung}.
One of the striking observations revealing the inherent topological character of the bands in graphene moire systems~\cite{senthilb,chittari} 
is the spontaneous anomalous Hall effects at 3/4 filling in tBG 
that appear even in the absence of an external magnetic field~\cite{goldhaber-gordon, lu2019superconductors, serlin2020intrinsic},
which can be viewed as a solid state realization of the Haldane-like model in a honeycomb lattice~\cite{haldane}.
Theories attempting to explain the spontaneous Hall effects relied on band structures 
that included a small mass term in the Dirac Hamiltonian contacting the BN substrate~\cite{zaletel,senthil} 
that give rise to precursor bands of the spontaneous quantum Hall phases.
The fact that this effect in tBG has been observed so far only in twisted bilayer graphene for samples that are nearly 
aligned with hexagonal boron nitride (tBG/BN) \cite{tschirhart2020imaging, polshyn2020nonvolatile, he2020giant}
motivates further research on the effects of the BN substrate in modifying the electronic structure of twisted bilayer graphene.

In this manuscript we show that commensurate double moire patterns of tBG on BN 
can have significant electron-hole asymmetric bands in addition to gaps between the moire bands
which provide the conditions for the formation of valley Chern bands prone to strong correlations.
%
%
% ---- modify below to highlight the new results ---- 
Here we focus our discussions on commensurate double moire patterns of tBG/BN 
whose continuum Hamiltonian can be conveniently formulated in terms in a unique moire Brillouin zone.
The electronic structure of double moire patterns was addressed in the past for arbitrary relative twist angles
in systems like BN encapsulated graphene~\cite{Nicolas2020a,Andelkovic2020aa,Wang2019aa} 
or twisted trilayer graphene~\cite{zhu2020twisted} giving rise to super-moire patterns. 
Earlier studies of BN/G/BN double moire systems~\cite{Nicolas2020a} have shown that the largest double moire 
interference effects are seen when the patterns are commensurate
and in these cases the electronic structure undergo significant modifications when interlayer sliding is introduced.

Electronic structure calculations for tBG/BN relying on commensurate double moire patterns
carried out recently have shown the opening of gaps between moire bands that
approximately maintains electron-hole symmetry~\cite{guinea, lin2020symmetry}.
%
% ---- ---- ----
%
In contrast, in our work we show that the moire bands can have significant electron-hole asymmetry 
depending on the relative sliding between the moire patterns and twist angles.
For type-1 double moire systems we can obtain on average a narrow conduction and a wider valence band,
while we predict that type-2 double moire systems can form ordered phases for both conduction and valence bands
 slightly favoring the valence bands.
Our results provides a possible justification that the spontaneously broken symmetry phases have been observed in experiments only for the conduction bands for the type-1 of double moire arrangements~\cite{goldhaber-gordon, andreayoung}, while correlated phases are likely for the valence bands for type-2.
In both cases we find that isolation of the flat bands required to endow finite valley Chern numbers 
is facilitated by the opening of primary and secondary gaps. 
We find that valley Chern numbers are finite with values $\pm 1$ 
over a large parameter space in keeping with experimentally observed quantized Hall conductivities, 
while the local density of states associated to the flat bands invariably concentrate around the AA-stacked 
regions of tBG regardless of the BN arrangement. 
We illustrate for a particularly pronounced electron-hole asymmetric band structure the impact of the
BN moire pattern by showing the differences in the optical 
conductivities between tBG and tBG/BN as a function of carrier density. 
Band structures corresponding to other rational numbers $N$-multiple commensurate moire periods 
indicates that it is possible to find narrow width $W \lesssim 10$~meV isolated folded 
band bundles for tBG angles $\theta \lesssim 1^{\circ}$.
% --- ---
%

The manuscript is structured as follows. In Sec.~II we introduce the model Hamiltonian for the 
doubly commensurate tBG on BN. In Sec.~III we present the exact geometric conditions under which 
the double moire patterns become commensurate with the same period and orientations.  
We then move on to analyze the electronic structures in Sec.~IV 
where we present the band structures for different relative sliding vectors
paying particular attention to the bandwidth, gaps, and valley Chern numbers,
and study the effects of interlayer potential differences due to a perpendicular electric field.
In Sec.~V we illustrate the effects of the BN substrate in the 
optical conductivity near the magic angle twisted bilayer graphene 
for a clearly electron-hole asymmetric electronic structure with a narrow conduction band. 
In Sec.~VI we discuss the electronic structures for commensurate double moire patterns
with different periods. 
Finally we close the paper in Sec.~VII with the conclusions.

%------------------------------------------------------------%
\begin{figure*}
\begin{center}
\includegraphics[width=0.98\textwidth]{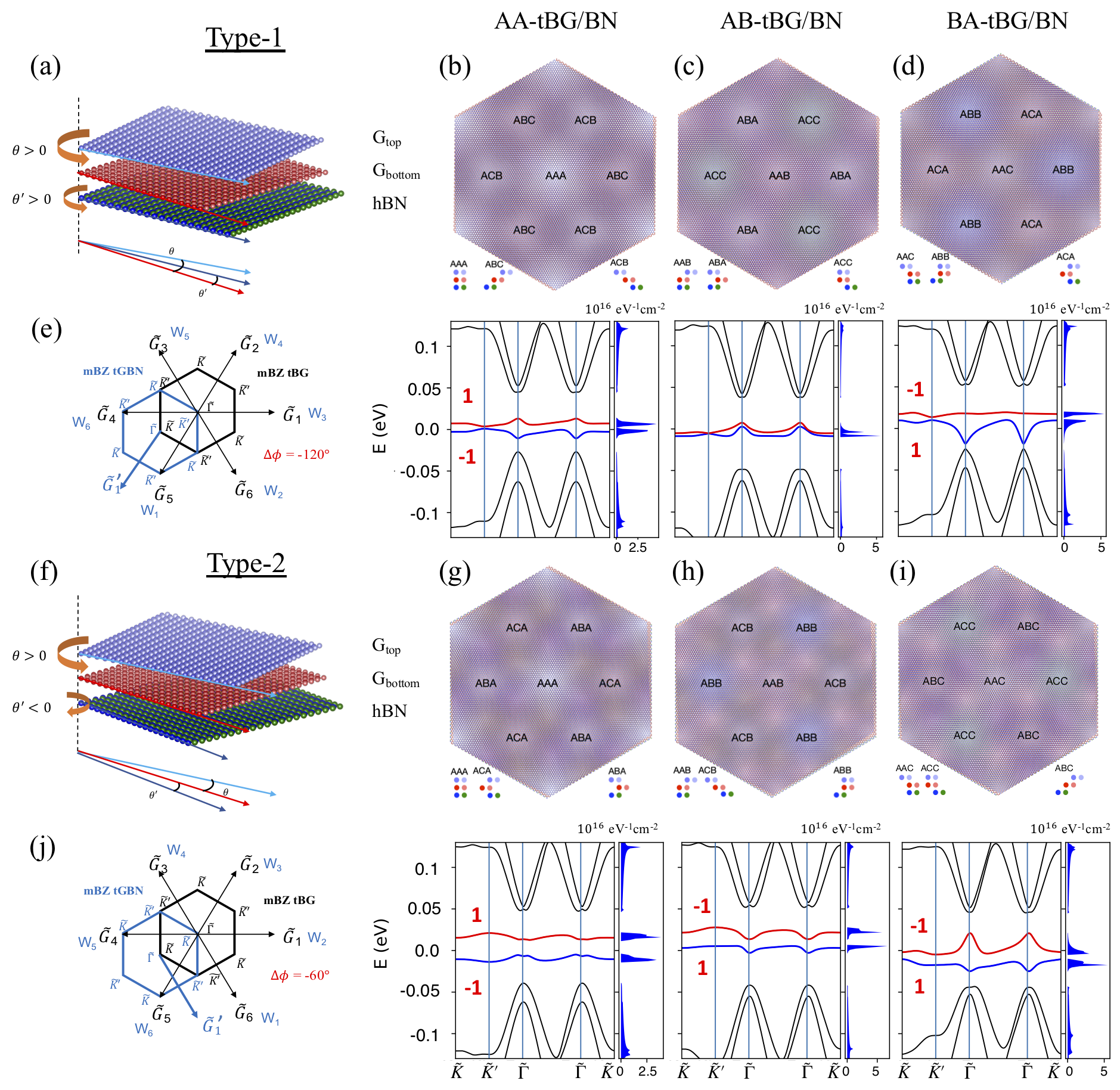}
\end{center}
\caption{
(Color online) Schematic diagrams of atomic configuration of tBG/BN for 
(a) type-1 ($\theta$, $\theta'$)=(1.13291$^\circ$, 0.5664$^\circ$) and 
(f) type-2 ($\theta$, $\theta'$)=(1.1463$^\circ$, $-0.5932^\circ$)
using the lattice constants of graphene $a_{\rm G} = 2.461~\AA$ and hexagonal boron nitride $a_{\rm BN} = 2.504~\AA$.
Moire Brillouin zone (mBZ) of tBG and tGBN for (e) type-1 and (j) type-2. Moire patterns and the corresponding energy bands in type-1 for (b) AA-tBG/BN, (c) AB-tBG/BN, (d) BA-tBG/BN, and in type-2 for (g) AA-tBG/BN, (h) AB-tBG/BN, (i) BA-tBG/BN. The first electron and hole bands are emphasized by red and blue colors, respectively. The distinct difference between type-1 and type-2 is in the direction of twist angle $\theta'$ between the bottom G layer and the hBN layer, which is positive for type-1 and negative for type-2. Underneath each moire pattern figure we illustrate three commensurate stackings with colored circles, blue and green for the hBN layer, red and tender red for the bottom G layer, and tender violet and tender blue for the top G layer. The labeling of stacking configuration starts from the lowest layer (hBN) and the sublattice on the left within a layer. Energy bands were plotted along the high-symmetry lines in (e) for type-1 and (j) for type-2. For type-1, mBZ of tBG and tGBN have an angle difference $\phi' - \phi = \Delta \phi = -120^\circ$, while $\Delta \phi = -60^\circ$ for type-2 (see Appendix A for more details). 
	}\label{fig1}
\end{figure*}

\section{Model Hamiltonian}\label{Model}
The model Hamiltonian for doubly commensurate tBG/BN can be conveniently formulated through
the moire bands theory with a single Brillouin zone~\cite{bistritzer11,Jung2014}
without needing to recourse to a supermoire framework for incommensurable double moire 
patterns~\cite{Andelkovic2020aa,moireofmoire,Wang2019aa,Nicolas2020a}.
The Hamiltonian matrix elements arising from the two moire interfaces from tBG and tGBN lead to two types of arrangements
illustrated in Fig.~\ref{fig1}(a)(f) that can be built by properly matching the orientation of the moire 
pattern angles with the moire reciprocal lattice vectors, see Fig.~\ref{fig1}(e)(j). 
The continuum model Hamiltonian is concisely given as $H = H_\textrm{\rm tBG} (\theta) + H_{\rm BN}^M(\bm{r})$, where $H_\textrm{\rm tBG} (\theta)$ is the Hamiltonian of tBG for the $K$ valley 
%\begin{equation}
%H_\textrm{\rm tBG} (\theta) = \left(\begin{array}{cc}
%h_b(-\theta/2) & T(\bm{r}) \\
%T^{\dagger}(\bm{r}) & h_t(\theta/2) \\
%\end{array}\right) + \frac{\eta}{2} \sigma_z,
%\end{equation}
\begin{equation}
H_\textrm{\rm tBG} (\theta) = \left(\begin{array}{cc}
h_b(-\theta/2) & T(\bm{r}) \\
T^{\dagger}(\bm{r}) & h_t(\theta/2) \\
\end{array}\right) + V,
\end{equation}
where $V$ = diag($\eta$, $\eta$, $-\eta$, $-\eta$)$/2$ represents the interlayer potential difference resulting from an
applied electric field in the perpendicular direction, and $H_{\rm BN}^M(\bm{r})$ describes the moire potential corresponding to BN placed underneath tBG, where $h_{b, t}(\pm \theta/2)$ represents a $2\times2$ Dirac Hamiltonian $v_F \bm{p} \cdot \bm{\sigma}$ for the bottom (b) and the top (t) layers of tBG rotated by $ \pm \theta/2$ ($+$: counterclockwise, $-$: clockwise), yielding
\begin{equation}
h_{b,t} (\pm \theta/2) = D(\mp \theta/2) ~ v_F \bm{p} \cdot \bm{\sigma} ~ D(\pm \theta/2),
\end{equation}
where $D(\pm \theta) = e^{\pm i \theta \sigma_z /2}$  and the Fermi velocity is $v_F = t_0\sqrt{3}a_G/2\hbar$, 
choosing $t_0 =3.1$ eV~\cite{Chebrolu2019}. 
Here, $\boldsymbol{\sigma} = (\sigma_x, \sigma_y)$, and $\sigma_z$ is the $z$-component of Pauli matrix. The interlayer coupling $T(\bm{r})$ is given by
\begin{equation}
T(\bm{r}) = \sum_{j=0, \pm} e^{-i \bm{q}_j \cdot \bm{r}} T^j_{s, s'},
\end{equation}
where $\bm{q}_0$, $\bm{q}_\pm$ are given as $\bm{q}_0 = \theta k_D (0, -1)$, 
$ \bm{q}_\pm = \theta k_{\rm D} (\pm \sqrt{3}/2, 1/2)$ if the twist angle $\theta$ is small enough. 
Here, $k_{\rm D} = 4\pi/3 a_{\rm G}$ is the magnitude of graphene Brillouin zone (BZ) corner 
wavevector where the lattice constant of graphene is given as $a_{\rm G} = 2.461 \textrm{\AA}$. 
The interlayer tunneling matrices $T^j_{s, s'}$ were formulated for the AA initial stacking 
in Ref.~\onlinecite{Jung2014} as 
\begin{equation}
%T^j_{s, s'} = \omega_{s,s'} \exp (ij \bm{G}_j \cdot (\bm{\tau}_s - \bm{\tau}_{s'})),
T^j =  e^{-i \bm{G}_j \cdot \bm{\tau}} \left(\begin{array}{cc}
\omega_{A,A'} & \omega_{A,B'} e^{-ij \varphi}  \\
\omega_{B,A'} e^{ij \varphi} & \omega_{B,B'} \\
\end{array}\right),
\label{eq_t}
\end{equation}
where $\bm{G}_0 = (0, 0)$ and $\bm{G}_\pm = k_D (-3/2, \pm \sqrt{3}/2)$ and $\tau = (\tau_x,~\tau_y)$ describes a relative sliding of the top G layer with respect to the bottom layer of graphene, and $\omega_{A,A'} = \omega_{B,B'} = \omega'$, and $\omega_{A,B'} = \omega_{B,A'} = \omega$, we get 
\begin{equation}
T^0 = \left(\begin{array}{cc}
\omega' & \omega\\
\omega & \omega' \\
\end{array}\right), \hspace{0.3 cm} T^{\pm} = \left(\begin{array}{cc}
\omega' & \omega e^{\mp i\varphi}\\
\omega e^{\pm i\varphi} & \omega' \\
\end{array}\right),
\end{equation}
for local AA-stacking corresponding to $\bm{\tau} = (0, 0)$, where $\varphi = 2\pi/3$. The local AB-stacking corresponds to $\bm{\tau} = a_G(0, 1/\sqrt{3})$, and the local BA-stacking is given by $\bm{\tau} = a_G(0, 2/\sqrt{3})$ or, equivalently, $\bm{\tau} = a_G(0, -1/\sqrt{3})$. By taking different $\omega' =  0.0939$~eV and $\omega = 0.12$~eV for the
intra- and inter-sublattice coupling between layers
related through $\omega' = A\omega^2 + B\omega + C$ where $A = -0.5506$, $B = 1.036$, and $C = -0.02245$ 
following the EXX+RPA total energy minima parametrization~\cite{Chebrolu2019,Park2020}. 

For tBG/BN, the moire potential Hamiltonian $H_{\rm BN}^M(\bm{r})$ has the form 
\begin{equation}
H_{\rm BN}^M(\bm{r}) = \sum_{m=1}^6 (W_{m} e^{i\bm{\tilde{G'}}_m \cdot \bm{r}} ) \oplus \bm{0}_{2 \times 2} \label{moireeq}
\end{equation}
where 
%\begin{widetext}
\begin{equation}
W_m = \left(\begin{array}{cc}
C_{\rm AA} e^{i[(-1)^{m} \phi_{\rm AA}]} & C_{\rm AB} e^{i[(-1)^{m} \phi_{\rm AB} - \varphi_m]} \\
-C_{\rm AB} e^{i[(-1)^{m} \phi_{\rm AB} + \varphi_m]}  & C_{\rm BB} e^{i[(-1)^{m} \phi_{\rm BB}]} \\
\end{array}\right). \label{gbn_moire}
\end{equation}
%\end{widetext} 
Here, $\varphi_m$ are the polar angles of six moire reciprocal lattice vectors $\tilde{\bm{G}}'_m$ of the tGBN interface 
with a prime symbol to distinguish from the unprimed moire reciprocal lattice vectors $\tilde{\bm{G}}_m$ for tBG. 
Here we use the moire potential parameter set given by $C_{\rm AA} = -14.88$ meV, $\phi_{\rm AA} = 50.19^\circ$, 
$C_{\rm BB} = 12.09$ meV, $\phi_{\rm BB} = -46.64^\circ$, $C_{\rm AB} = 11.34$ meV, 
$\phi_{\rm AB} = 19.60^\circ$ following Refs.~\onlinecite{Jung2015aa,Jung2014}.

For tBG, the reciprocal lattice vectors of BZ of graphene are $\bm{G}_1 = 2\pi (0, 2/\sqrt{3}a_{\rm G})$, $\bm{G}_m = \hat{R}_{\pi(m-1)/3} \bm{G}_1$ and the moire reciprocal lattice vectors are given by $\bm{\tilde{G}}_i = - \theta \hat{z} \times \bm{G}_i$ ($i = 1, \cdots , 6$). On the other hand, for tGBN we have $\bm{\tilde{G'}}_i = \epsilon \bm{G}_i - \theta \hat{z} \times \bm{G}_i$. We assume that for the commensurate angle sets (see the next section for further details), the mBZ of tBG and tGBN have the same size and share their two corners in momentum-space as shown in Fig.~\ref{fig1} (e), (j) (see Appendix A for further details).

We label each commensurate stacking in real-space with three characters, A, B, and C to label 
from bottom to top layers the relative position of the left sublattice of the two atoms unit cell of G or BN. 
For instance, when the local AA-stacking region in tBG is AA stacked on top of hBN (AA-tBG/BN), 
a type-1 system gives rise to three AAA, ABC, and ACB local stacking at the symmetry points,
while AAA, ACA, and ABA local stacking appear in type-2 systems. 
We plot each of the commensurate stacking configurations of tBG/BN with six small colored 
circles under the moire patterns for type-1 in Fig. \ref{fig1} (b)-(d) and for type-2 in (g)-(i).

\begin{figure*}
\begin{center}
\includegraphics[width=1.0\textwidth]{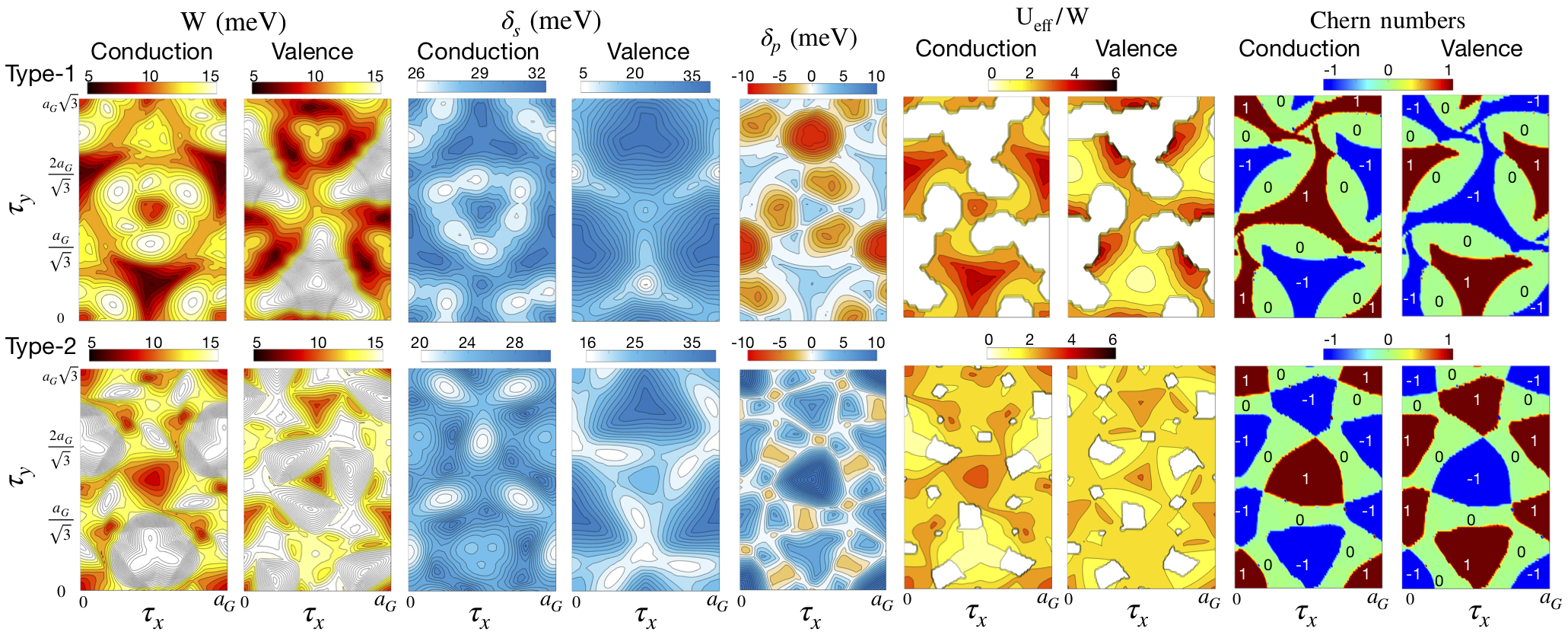}
\end{center}
\caption{
	(Color online) Bandwidth W, secondary $\delta_s$ and primary $\delta_p$ gaps, ratio of the effective Coulomb interaction to W, and valley Chern number of conduction and valence band as a function of sliding vector of graphene on graphene $\bm{\tau} =(\tau_x, \tau_y)$ for type-1 (upper row) and type-2 (lower row) systems. The commensurate stacking with a translation by $\tau_y = a_G/\sqrt{3}$  and $\tau_y = 2a_G/\sqrt{3}$ along the line $\tau_x = 0$ represents the local-AB and local-BA stackings in tBG respectively.
	}\label{fig_bw_surf}
\end{figure*}

%------------------------------------------------------------%
\section{commensurate twist angle sets}\label{comm_twist_ang}
Among the many possible commensurate double moire patterns~\cite{Nicolas2020a,Andelkovic2020aa,Wang2019aa} 
here we want to focus on the subset that can be obtained by requiring two constraints: (1) moire length: the moire lengths of tBG and tGBN should be the same, and (2) moire angle: the difference of moire angle should be 0 (mod $60^\circ$). The moire length of tBG is
\begin{equation}
L^M_\textrm{tBG} = \frac{a_{\rm G}}{2 \sin (\theta/2)}, \label{moirelen1}
\end{equation}
and the moire length of tGBN is 
\begin{equation}
L^M_\textrm{tGBN} = \frac{a_{\rm G}}{\sqrt{\alpha^2 - 2\alpha\cos \theta' +1 }},  \label{moirelen2}
\end{equation}
where $\alpha = a_{\rm G}/a_{\rm BN} =  1 + \epsilon$ and 
$\epsilon = (a_{\rm G} - a_{\rm BN})/a_{\rm BN}$ \cite{Jung2014} are parameters
that represent the lattice mismatch between two layers. 
We use for the lattice constant of graphene $a_{\rm G} = 2.461~\AA$ 
and boron nitride $a_{\rm BN} = 2.504~\AA$, which results in $\epsilon = -0.017172$.
Here, we use the notation for angles measured with respect to the bottom graphene (G) layer for 
convenience since both moire interfaces are sharing it; 
we define $\theta = \theta_{\rm tBG}$ as the twist angle of tBG which is measured with 
respect to the bottom G layer and $\theta' = -\theta_{\rm tGBN}$ is the twist angle of 
BN measured with respect to the contacting bottom G layer at the tGBN interface and hence the minus sign.
The condition for coincident moire lengths 
$L^{M}_\textrm{tBG} = L^{M}_\textrm{tGBN}$ requires the following
relation between $\theta$ and $\theta' $
\begin{equation}
\theta = \cos^{-1} \bigg[ \frac{(1+\alpha)(1-\alpha) + 2 \alpha \cos \theta'}{2}  \bigg]. \label{relation_th_thp}
\end{equation}
%\theta = \cos^{-1} \bigg[ \frac{(\alpha+1)(\alpha-1) + 2 \alpha \cos \theta'}{2 \alpha^2}  \bigg].

Accordingly, the rotation angle $\phi = \phi_{\rm tBG}$ of the moire pattern in tBG with respect to the bottom G layer is~\cite{Jung2014}
\begin{eqnarray}
\phi = 
%\phi_{G/G} = 
\tan^{-1} \left( \frac{\sin \theta}{\cos\theta - 1} \right).
\end{eqnarray}
This expression reduces to 
\begin{eqnarray}
\phi &=& - {\rm sgn}{(\theta)} \frac{\pi}{2} +  \frac{\theta}{ 2 }
\end{eqnarray}
indicating that the moire pattern orientation rotates with respect to the bottom G %with respect to the bottom tGBN moire
by one half of the total twist angle between the graphene layers. 
%
% phiGBN
%
On the other hand, the tGBN moire pattern rotation angle $\phi' = \phi_{\rm tGBN}$ measured with 
respect to the bottom G layer is given by 
\begin{eqnarray}
\label{comm1}
\phi^{\prime} 
% = \phi_{G/BN} %&=& \tan^{-1} \left( \frac{\alpha \sin( \theta_{G/BN})}{\alpha \cos( \theta_{G/BN}) - 1} \right) \\
 & = &  \tan^{-1} \left(  \frac{ \alpha \sin \theta'}{\alpha \cos\theta' - 1} \right).
\end{eqnarray}
%The twist angle $\theta_{G/BN}$, the twist angle of bottom G with respect to BN.
% $\theta^{\prime} = -\theta_{G/BN}$ is the rotation angle of BN measured with respect to the contacting graphene layer and  % contacting bottom BN.
%
%
The condition for the two equal period moire patterns to be commensurate is 
\begin{eqnarray}
\Delta \phi = \phi^{\prime} - \phi =  \frac{n \pi}{3} 
\label{anglecomm}
\end{eqnarray}
where $n$ is an integer number \cite{Nicolas2020a}. There is no analytical solution for $n=0$ but we have for $n = 1$ the type-1 solution at $\Delta \phi = 60^\circ$ or equivalently, $-120^\circ$ because of $\pi$ periodicity of the tangent function, 
and have for $n = 2$ the type-2 solution at $\Delta \phi = -60^\circ$ or equivalently $120^\circ$.
The rotation of the moire patterns and associated moire Brillouin zones are shown graphically in Fig.~\ref{fig1} (a), (f) in real-space and Fig.~\ref{fig1} (e), (j), and Fig.~\ref{Afig1} in appendix A in momentum space. The main differences between the two solutions are that in type-1 systems the top G layer and bottom hBN layer are rotated in the same sense $e.g.\,\,(\theta>0$, $\theta'>0$) with respect to the bottom G layer, while in type-2 they are rotated in the opposite directions with respect to the bottom G layer $e.g.\,\,(\theta>0$, $\theta'<0$). 
Imposing equal moire length $L_{\rm tBG}^{M} = L_{\rm tG/BM}^M$ conditions from Eqs.~(\ref{moirelen1},\ref{moirelen2}) leads to 
\begin{eqnarray}
%\pm 4 \sin^{2} (\theta / 2) =   (\alpha-1)^2 + 2 \alpha (1 - \cos (\theta')) 
4 \sin^{2} (\theta / 2) =  (\alpha- \cos (\theta^{\prime}))^2 + \sin^2(\theta'),
\label{mlength}
\end{eqnarray}
that is equivalent to Eq.~(\ref{anglecomm}) with $n=1$ when we use the $\theta = 2 \theta^{\prime}$ relation~\cite{guinea,shi2020moir} reducing to
\begin{eqnarray}
\alpha  \pm \sqrt{3} \sin(\theta^{\prime})  =  \cos(\theta^{\prime}),
\end{eqnarray}
that yields same sign angles for type-1 solutions 
$(\theta, \theta') = (1.1329^{\circ}$, $0.5664^{\circ})$.
%
%The moire period of G/G moire pattern is given by 
%\begin{eqnarray}
%L_{G/G} = a / 2 \sin (\theta / 2) 
%\end{eqnarray}
%and the moire pattern of BN is given by
%\begin{eqnarray}
%L_{G/BN} = \alpha a_{BN} /\sqrt {( (\alpha-1)^2 + 2 \alpha (1 - \cos (\theta')) )}
%\end{eqnarray}
%which reduces to the condition
%\begin{eqnarray}
%\pm 4 \sin^{2} (\theta / 2) =  (\alpha-1)^2 + 2 \alpha (1 - \cos (\theta')) 
%\end{eqnarray}
%
The equations for unequal sign angles for type-2 solutions can be formulated by combining both Eqs.~(\ref{anglecomm},\ref{mlength})
for $n=2$ and their numerical solutions are $(\theta, ~\theta') = (1.1463^{\circ}$, $-0.5932^{\circ})$. 
Geometric discussions of the double moire commensuration conditions in real and momentum space are discussed in Appendix A. 
%,$\theta = 1.14^{\circ}$ and $\theta^{\prime} = -0.59^{\circ}$.
%
%The commensurate angle sets for type-1 and type-2 systems are and $(1.1463^{\circ}$, $-0.5932^{\circ})$, respectively.  
The real-space moire patterns of the commensurate double moire patterns for the 
three different stackings configurations between the top and bottom G layers 
for both type-1 in Fig.~\ref{fig1} (b-d) and type-2 systems are shown in Fig.~\ref{fig1} (g-i) 
where we can clearly see their differences both in the real-space patterns
as well as in their momentum space band structures.

%------------------------------------------------------------%

\section{Band structures}\label{R1}
In this section we present the band structures of tBG/BN corresponding to the two types of commensurate stacking arrangements introduced in earlier sections as a function of moire pattern sliding and perpendicular external fields.
We will be paying attention to the bandwidth, gaps, and electron-hole asymmetry.
Knowledge of the band structure %width $W$ and the $\delta_p$, $\delta_s$ gaps 
allows to estimate regimes of strong correlations that satisfy 
$U_\textrm{eff}/W \gtrsim 1$ where the bandwidths $W$ are sufficiently narrow 
versus the effective Coulomb interaction %$U_\textrm{eff}$ 
\begin{equation}
U_\textrm{eff} = \frac{e^2}{4 \pi \epsilon_r \epsilon_0 l_M} \exp{(-l_M/ \lambda_D)}
\end{equation}
where we used the dielectric constant of graphene as $\epsilon_r = 4$, 
the moire length $l_M = a_G/(2 \sin(\theta/2)) \simeq  a_{G}/\theta $ and
the screening length $\lambda_D = 2 \epsilon_0 / e^2 D(\delta_p, \delta_s)$ that depends on 
the two-dimensional density of states
$D(\delta_p, \delta_s) = 4 ( \vert \delta_p \vert u(-\delta_p) + \vert \delta_s \vert u(-\delta_s) )/ (W^2 A_M )$
defined in terms of the Heaviside step function $u(x)$ that includes screening due to overlap between neighboring bands.~\cite{Chebrolu2019} 
The valley Chern number of the $n^{\rm th}$ energy band 
$C_n =  \int_\textrm{mBZ} d^2 \bm{k} ~ \Omega_n (\bm{k}) / (2 \pi)$
is obtained integrating the Berry curvature~\cite{Xiao2010b} given by
$\Omega_n (\bm{k}) = -2 \sum_{n' \neq n} \textrm{Im} \left[ {\langle n \vert \frac{\partial H}{\partial k_x} \vert n' \rangle \langle n' \vert \frac {\partial H}{\partial k_y} \vert n \rangle}{/(E_{n'} - E_n)^2} \right]$.
Additional information on the electronic structure can be found in appendix C where we present 
the normalized local density of states (LDOS) whose maxima are expected at the AA stacking sites of tBG.

\subsection{Type-1 and type-2 electronic structures}\label{R2}
We can distinguish two arrangements of double moire patterns that we classify as type-1 and type-2
depending on the relative twist of the layers in the tBG and tGBN substrate that in turn impacts
the relative rotation of the associated moire patterns $\Delta \phi = \phi' - \phi$.  %\phi_{\rm tBG} - \phi_{\rm tGBN} $. 
For double moire patterns of type-1 
we require a relative rotation of $\Delta \phi  = -120^\circ$ or equivalently $\Delta \phi = 60^\circ$
between the mBZ of tBG and tGBN as shown in Fig. \ref{fig1}(e). 
On the other hand, for type-2 we require
$\Delta \phi = -60^\circ$ (see Appendix A) or equivalently $\Delta \phi = 120^\circ$ as shown in Fig.~\ref{fig1}(j). 
The impact in the band structure of  tBG/BN
is represented along the high-symmetry line of mBZ of tBG ($\tilde{K}-\tilde{K}'-\tilde{\Gamma}-\tilde{\Gamma}-\tilde{K}$), 
we show the band structures for in Fig. \ref{fig1}(b-d) for type-1 and in Fig.~\ref{fig1} 
(g-i) for type-2 for different displacement of the top G layer for an electronic structure model 
with unequal interlayer tunneling value that incorporates an effective vertical relaxation~\cite{Chebrolu2019,Park2020}. 
In our analysis the twist angles of tBG compatible with the double commensuration defined by 
Eqs.~(\ref{comm1})-(\ref{anglecomm}) essentially depends on the value of the lattice constant 
mismatch $\alpha = a_{\rm G} / a_{\rm BN}$.
The double commensuration angles $\theta \simeq 1.13^{\circ}$ and $\theta \simeq 1.14^{\circ}$ result 
from our choice of  graphene and BN lattice constants.
We can expect the bandwidths will remain narrow provided that these angles remain close to the 
experimental~\cite{cao2020nematicity} $\theta^{\rm exp}_{\rm MA} \simeq 1.08^{\circ}$
and theoretical~\cite{bistritzer11,Chebrolu2019,Park2020} $\theta^{\rm th}_{\rm MA} \simeq 1.05^{\circ}$ values where flat bands emerge.
We estimated the changes in the electronic structure for twist angles $\theta$ of tBG away from the magic angle 
by using an approximation scheme that preserves commensuration with the hBN moire pattern potential which 
shows a similar qualitative behavior in the vicinity of the magic twist angle
for variations $\delta \theta$ on the order of $\sim 10\%$ or smaller, see appendix C Fig.~\ref{fig_bw_ueff_cnum}.

The electronic structures depend strongly on the relative sliding between the moire patterns of tBG and tGBN.
We use a vector $\bm{\tau} = (\tau_x,~\tau_y)$ to indicate the sliding of the top G layer prior to rotation and with this notation the local AA-, AB-, and BA-stackings at the origin 
correspond to $(\tau_x, ~\tau_y) = (0, ~0)$, $a_G(0,~1/\sqrt{3})$, and $a_G(0,~2/\sqrt{3})$. 
The local stacking maps between the two graphene sheets and BN subsrate are traced by
assuming that the BN substrate is locked at the AA stacking configuration with respect to the bottom graphene layer. 
For instance, the AA-tBG/BN case where tBG is AA stacked prior to twisting corresponds to Fig.~\ref{fig1}(b) and (g).
For type-1 we can find a small gap of $\sim 2.5$~meV at $\tilde{K}'$
while for type-2 the lowest energy bands are quasi-flat with a large gap of 
$\sim 35$~meV at $\tilde{K}'$ and $\sim 20$ meV at $\tilde{\Gamma}$. 
For AB-tBG/BN corresponding to Fig.~\ref{fig1}(c) and (h), type-1 bands consist of mutually overlapping
bands with large secondary gaps $\delta_s$ for both conduction and valence bands that separate the lowest energy 
bands with the next band that is farther in energy from charge neutrality, 
whereas for type-2 we find isolated low energy conduction and valence bands with a $\delta_p$ gap of 
size $\sim 15$~meV at $\tilde{\Gamma}$. 
For BA-tBG/BN corresponding to Fig.~\ref{fig1} (d), we see a large electron-hole asymmetric 
electronic structure separated with a small primary gap $\delta_p \simeq  3.5$~meV 
consisting of a nearly flat low energy conduction band with $W \sim 5$~meV 
and wider valence band $W \sim 30$~meV that reminds of the experimental observations of 
spontaneous Hall effects for electrons but not for holes~\cite{goldhaber-gordon, andreayoung}. 
Conversely for type-2 we have wider conduction and narrower valence bands
that are separated by a finite gap of $\delta_p \simeq 5.6$~meV for the entire mBZ. 

For both types of solutions, the secondary gap $\delta_s$ for the conduction and valence bands remain open 
for the three AA, AB, BA symmetric stacking geometries, 
and we will show shortly that this is the case for the entire range of sliding geometries. 
This is in contrast to the behavior of the primary gap $\delta_p$ that clearly closes when the top G layer is placed 
at $(\tau_x,~ \tau_y) = a_G(0,~1/\sqrt{3})$ in a type-1 system.
%

% ---- LDOS and Berry curvatures ----

The Berry curvatures for discrete three commensurate slidings of the top G layer for conduction and valence band and the normalized 
local density of states $\tilde{D}(r, E) = D(r, E)/{\rm max} (\vert D(r, E) \vert)$ at the van Hove singularity 
(vHs) can be found in appendix C Fig.~\ref{fig_ldos_tau} that illustrates the impact of the BN substrate
in altering those quantities. 

% --- --- --- --- --- --- --- --- --- --- --- --- ---

Having identified the distinct behavior of the bands for type-1 and type-2 moire pattern arrangements
at three different symmetric stacking configurations we discuss in the following the dependence of the 
band structure to other continuous stacking sliding geometries and interlayer potential differences.

\begin{figure*}
\begin{center}
\includegraphics[width=1.0\textwidth]{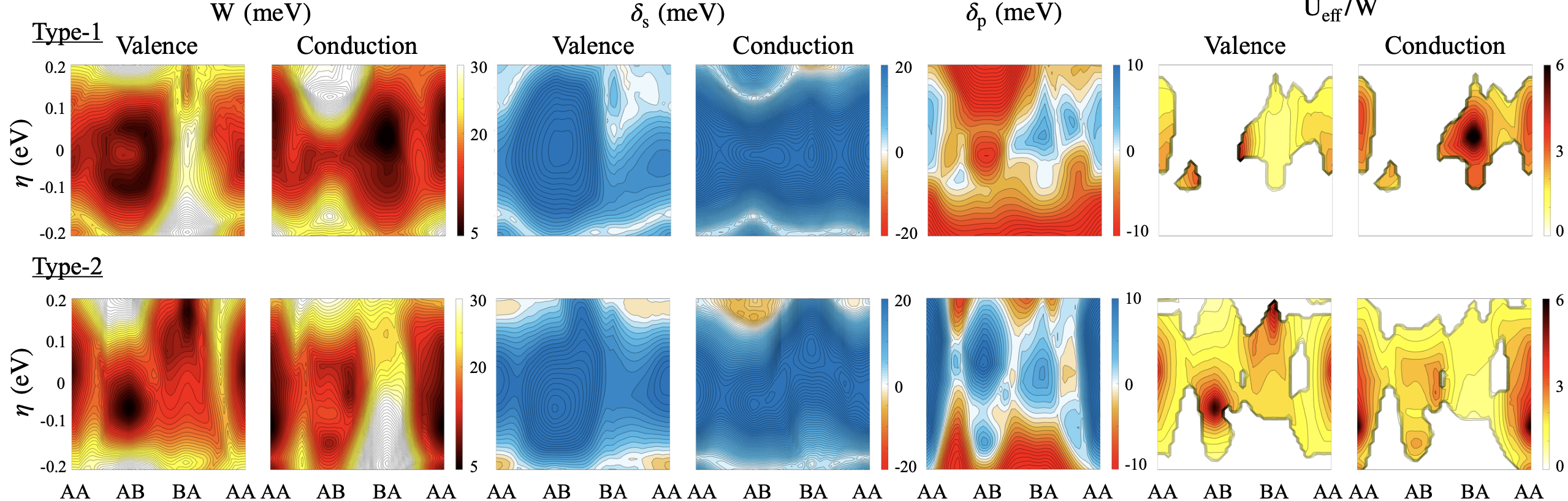}
\end{center}
\caption{
	(Color online) Bandwidth W, secondary $\delta_s$ and primary $\delta_p$ gaps, ratio of the effective Coulomb interaction to W for type-1 (upper row) and type-2 (lower row) 
	as a function of continuous sliding of the top G layer in the $y$-direction ($\tau_y$) prior rotation obtained for various values of the interlayer potential difference $\eta$. We see that both stacking configuration and electric fields can substantially alter the bandwidth and band isolation of the low energy moire bands. 
	}\label{efield}
\end{figure*}

\subsection{Sliding and electric field dependent electronic structure}\label{R3}   
Here we extend our analysis of the lowest energy bands to other continuous stacking geometries as a function 
of $\bm{\tau}=(\tau_x, \tau_y)$ and explore the changes in the electronic structure due to an interlayer potential difference $\eta$
introduced by a perpendicular electric field. 
\begin{table}[]
\begin{tabular}{|c|r|r|r|r|}
\hline
\multicolumn{1}{|c|}{} & \multicolumn{2}{c|}{ Type-1} & \multicolumn{2}{c|}{Type-2}  \\ \hline
 $W$ (meV) & {$\rm cond.$} & {$\rm val.$}  & {$\rm cond.$} & {$\rm val.$}      \\ \hline 
 $\rm max$  & 15.2  & 29.2 &  26.5  & 21.7    \\ \hline
 $\rm min$   &   6.0  &   6.0 &    8.2  &   8.9    \\ \hline
  $\rm av$    & 11.6  &  13.1 & 15.0  &  14.8   \\ \hline 
  $\rm std$   &   1.9  &   4.6 &   4.8   &    2.6   \\ \hline 
\end{tabular}
\caption{Bandwidth $W$ maxima, minima, and average taken from every possible local sliding configurations 
and the corresponding standard deviation for type-1 and type-2 moire pattern stacking arrangements. 
We find narrower bandwidths with smaller fluctuations for the type-1 conduction and type-2 valence bands.}
\label{bands_average}
\end{table}
For all possible local stacking geometries we have obtained the maxima, minima, 
average and the standard deviations of the bandwidths, 
which we succinctly summarize in Table~\ref{bands_average}.
From this data we can observe a more pronounced electron-hole asymmetry for type-1 systems 
with narrower conduction bands on average and smaller bandwidth fluctuations that posits the preference 
of the conduction bands to form ordered phases with respect to the valence bands. 
On the other hand, we expect a higher likelihood of observing ordered phases for both conduction and 
valence bands for type-2 systems because the electron-hole asymmetry is generally less pronounced, 
while narrower valence bands with smaller fluctuations are found for the average results.

The phase diagram maps for the bandwidth $W$, the $U_{\rm eff}/W$ ratios, 
the  $\delta_p, \, \delta_s$ gaps, and valley Chern number maps for continuous stacking geometries are shown in Fig.~\ref{fig_bw_surf}.
The stacking configurations with the narrowest bandwidth and largest gaps 
have greater chances of enhanced Coulomb interactions that can lead to ordered phases.
For roughly half of the stacking configuration space we can observe favorable 
$U_{\rm eff}/W \gtrsim 1$ condition and they largely coincide with the regions 
that have finite valued valley Chern numbers. 
The regions with suppressed $U_{\rm eff}/W$ are mainly due to the closure of the primary 
gap $\delta_p$ that overlaps the two low energy flat bands and enhances the Coulomb screening.

%Our calculations show that a perpendicular electric field directed in the appropriate sense can enhance the narrowing of the bands and has an overall effect of enhancing the secondary gap $\delta_s$, while the primary gap $\delta_p$ follows a more complex behavior as a function of stacking and applied electric fields. 

The phase diagram maps for the bandwidth $W$, the gaps and resulting $U_{\rm eff}/W$ ratios 
in Fig.~\ref{efield} shows that a perpendicular electric field directed in the appropriate sense can enhance the 
narrowing of the bands and has an overall effect of enhancing the secondary gap $\delta_s$, 
while the opening and closure of the primary gap $\delta_p$ follows a more complex behavior 
as a function of stacking and applied electric fields.

\subsection{Band gap analysis at $\tilde{K}'$ point} \label{R3}
As discussed earlier, distinct electronic low energy bands are expected depending 
on the moire pattern stacking types that we classify as type-1 and type-2. 
For example, in type-1 of AA-tBG/BN there is a small gap $\sim 2.5$~meV 
at $\tilde{K'}$ in contrast to the large gap $\sim 35$~meV in type-2. 
A band gap near charge neutrality in tBG/BN double moire systems can be expected to appear 
due to moire pattern interference effects. 
We illustrate this behavior by obtaining the analytical expressions for the size of the 
gaps at $\tilde{K}'$ for different moire pattern types and relative stackings arrangements
as a function of the model parameters that define the moire patterns.
For this purpose we consider a truncated $8\times8$ Hamiltonian as a simple model to describe tBG system 
that captures the scattering of the electrons from the three Dirac cones of the bottom layer to one Dirac cone on the top layer~\cite{bistritzer11}. 
At $\tilde{K}'$ the eigenstates can be represented by $\Psi = (\psi_1, \psi_2, \psi_3, \psi_4)^T$ consisting of 
four two-component sublattice pseudospin spinors~\cite{Javvaji2020}. Here the $\psi_2$, $\psi_3$ spinors can be rewritten in 
terms of $\psi_4$ by multiplying with a unitary matrix $\psi_2 = U \psi_4$ and $\psi_3= U^\dagger \psi_4$, 
and thus allowing to further reduce our Hamiltonian to a 3$\times$3 matrix to obtain a simpler characteristic equation. 
Then, the expression for the band gap at $\tilde{K}'$ is,
\begin{equation}
E_{C}(\tilde{K}') - E_{V} (\tilde{K}') \approx \frac{6~(A_3 - B_3)}{3\omega'^2 +3\omega^2+ Q^2}, 
 \label{analyticGap}
\end{equation}
where $Q = \hbar \upsilon_F \theta k_D$. 
This expression agrees up to 2 significant digits with the gap size obtained by diagonalizing numerically the eight-band model. The above equation holds for both type-1 and type-2 solutions and for the three symmetric AA, AB, BA stackings but the analytical coefficients $A_3$ and $B_3$ are different for each case.
Below we present the expression for the $A_3$ and $B_3$ coefficients for type-1 BA-tBG/BN for the electronic 
structure showing the largest electron-hole asymmetry that incorporates the effect of the interlayer 
potential difference $\eta$ that can be induced by a perpendicular electric field and the moire pattern potentials
generated by the BN substrate that affects the magnitude of the gap. 
For sufficiently small magnitudes of $\eta$ the $A_3$ and $B_3$ coefficients in Eq.~\ref{analyticGap} can
be expressed in terms of the moire pattern coefficients stemming from the hBN substrate as follows
\begin{eqnarray}
%\begin{array}{ll}
A_3 &= &\omega'^2 C_{\rm AA}\cos(\phi_{\rm AA}+\varphi) + \omega^2 C_{\rm BB}\cos(\phi_{\rm BB}-\varphi) \\
&+&\omega'\omega C_{\rm AB} \cos(\phi_{\rm AB}-\varphi) \nonumber \\
&-& ( \eta/6 ) \big(-Q^2 +4C_{\rm AA}{\rm cos}(\phi_{\rm AA}+\varphi) C_{\rm BB}{\rm cos}(\phi_{\rm BB}-\varphi) \nonumber \\
&-& 4C_{\rm AB}^2 \cos^2(\phi_{\rm AB}-\varphi) -2\sqrt{3}QC_{\rm AB} {\rm cos}(\phi_{\rm AB}-\varphi) \big), \nonumber
%\end{array}
\end{eqnarray}
and 
\begin{eqnarray}
%\begin{array}{ll}
B_3&=& \omega^2 C_{\rm AA}\cos(\phi_{\rm AA}-\varphi) + \omega'^2 C_{\rm BB}\cos(\phi_{\rm BB}) \\
&+& \omega'\omega C_{\rm AB}\cos(\phi_{\rm AB}+\varphi)) \nonumber \\
&-& ( \eta/6 ) \big( -Q^2 + 4C_{\rm AA}\cos(\phi_{\rm AA}-\varphi) C_{\rm BB}\cos(\phi_{\rm BB}) \nonumber \\
&-& 4C_{\rm AB}^2\cos^2(\phi_{\rm AB}+\varphi) -2\sqrt{3}QC_{\rm AB}\cos(\phi_{\rm AB}+\varphi) \big). \nonumber
%\end{array}
\end{eqnarray}
In Appendix B we show a more detailed derivation of 
the analytic expressions for other types of solutions and stackings.

\section{Optical conductivity}
\begin{figure}
\begin{center}
\includegraphics[width=0.48\textwidth]{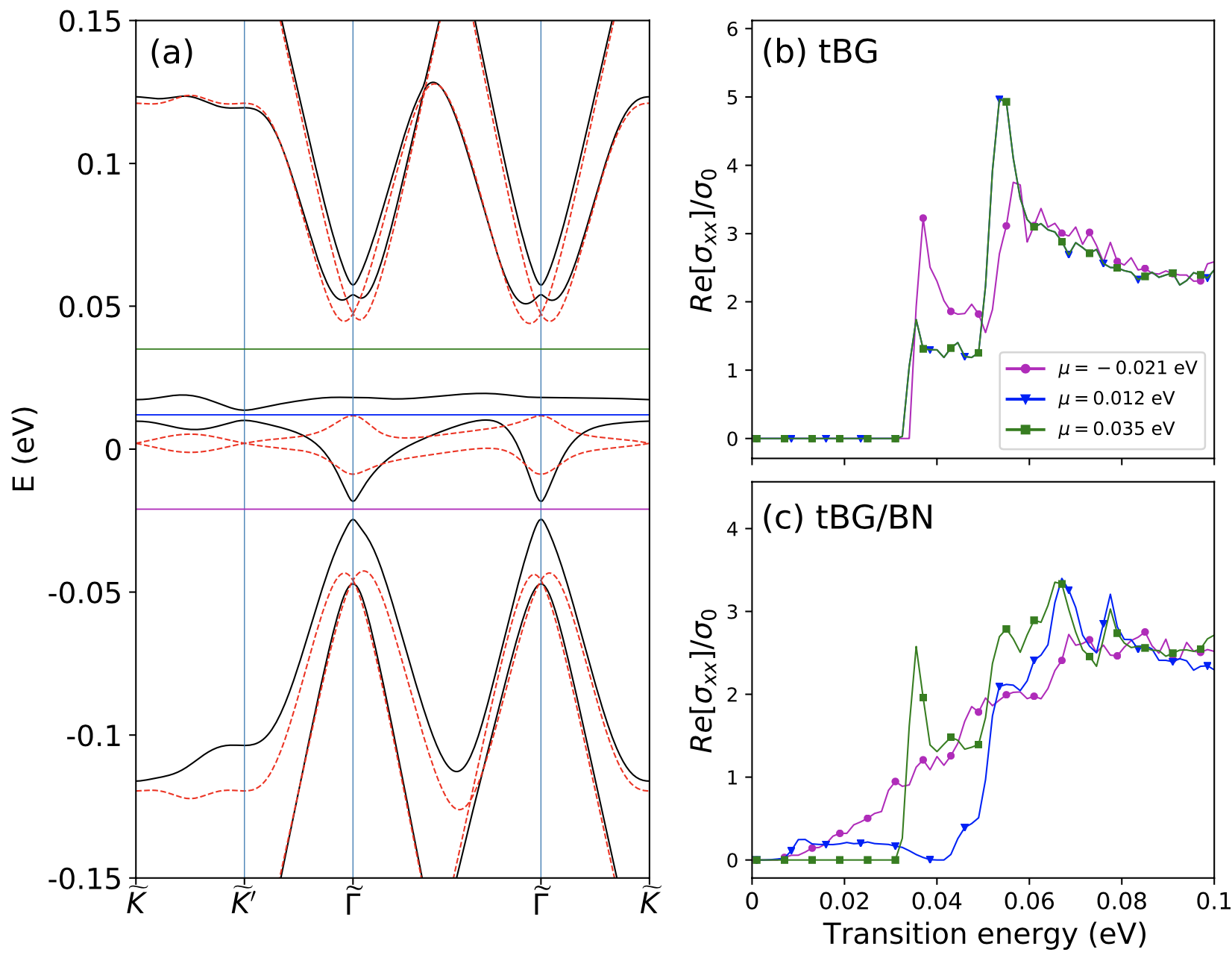}
\end{center}
\caption{
	(Color online) (a) Energy bands under the effective vertical relaxation for BA-tBG/BN (black solid line) in type-1 and for tBG (red dashed line) for comparison. 
	The horizontal magenta, blue and green lines represent the positions of the Fermi energy that empties or 
	fills a nearly flat band. 
	Longitudinal optical conductivity in (b) tBG, and in (c) tBG/BN in the case of the set of twist angles ($\theta$, $\theta'$) = (1.13291$^\circ$, 0.5664$^\circ$) for different three values of chemical potential $\mu = -0.021$~eV (magenta circle), $0.012$~eV (blue triangle), and $0.035$~eV (green square) corresponding to the horizontal lines in panel (a). 	}\label{fig7_opt}
\end{figure}

Electronic structure reconstruction in the energy ranges of a few tens of meV 
in the vicinity of the Dirac cone of graphene can be conveniently explored by means of mid-infrared 
and terahertz spectroscopy. 
Here we show the numerical results for the optical absorption in type-1 BA-tBG/BN as well as tBG without a BN 
substrate in an effort to estimate the differences in the optical signals that can be expected in these two systems. 
The real part of the longitudinal optical conductivity normalized by the universal optical conductivity of graphene$\sigma_0 = \pi e^2/2h$ is given by~\cite{ando2002dynamical, gusynin2006unusual, gusynin2007anomalous, falkovsky2007space, min2009}
\begin{widetext}
\begin{equation}
Re[\sigma_{xx}(\omega)]/\sigma_0 = \frac{16}{\omega} \int \frac{d^2 \bm{k}}{(2\pi)^2}\sum_{i,j} [f(\epsilon_{\bm{k},i})-f(\epsilon_{\bm{k}, j})] \vert \langle \bm{k}, i \vert  J_x \vert \bm{k}, j\rangle \vert^2 \delta[\omega + (\epsilon_{\bm{k}, j} - \epsilon_{\bm{k},i})/\hbar],
\end{equation}
\end{widetext}
where $J_\alpha = - \partial H / \partial k_\alpha$ is the current operator, $f(\epsilon)$ is the Fermi-Dirac distribution function. 
$\epsilon_{\bm{k}, i}$ is the $i$th eigenstate energy at $\bm{k} = (k_x,~ k_y)$. 
In Fig.~\ref{fig7_opt}(a) we show the energy bands for type-1 BA-tBG/BN (black solid line) and 
tBG (red dotted line) without a BN substrate where we achieve double commensuration of equal period moire patterns
by using the twist angles of ($\theta$, $\theta'$) = (1.13291$^\circ$, 0.5664$^\circ$) as we introduced earlier. 
The horizontal lines are the three different chemical potentials $\mu =-0.021$ (magenta circle), 0.012 (blue triangle), and 0.035 eV (green square) considered, and we plot the real part of the linear optical absorption in the longitudinal direction for tBG in Fig.~\ref{fig7_opt}(b) and type-1 BA-tBG/BN in Fig.~\ref{fig7_opt}(c). 
The presence of the BN substrate leads to markedly different asymmetric optical absorptions for $\mu =-0.021$~eV (magenta circle) and $0.035$~eV(green square), and one can observe small gaps less than 10~meV for $\mu =-0.021$~eV (magenta circle), $0.012$~eV (blue triangle) in type-1 BA-tBG/BN as shown in Fig.~\ref{fig7_opt}(c), unlike without BN.

\begin{table*}[]
\begin{tabular}{|c|c|r|r|r|r|r|r|r|}
\hline
\multicolumn{9}{|c|}{$L^M_{tGBN}$ = $N$ $\times L^M_{tBG}$} \\
\hline
\multicolumn{2}{|c|}{$N$} & \multicolumn{1}{c|}{$1/4$} & \multicolumn{1}{c|}{$1/3$} & \multicolumn{1}{c|}{$1/2$} & \multicolumn{1}{c|}{$1$} & \multicolumn{1}{c|}{$2$} & \multicolumn{1}{c|}{$3$} & \multicolumn{1}{c|}{$4$} \\ \hline
\multirow{2}{*}{type-1} & $\theta^{\circ}$ & 0.284447 & 0.379079 & 0.568072 & 1.132910 & 2.253321 & 3.362061 & 4.459900 \\ \cline{2-9} 
 & $\theta^{\prime \circ}$ & 0.576353 & 0.575245 & 0.573035 & 0.566455 & 0.553515 & 0.540851 & 0.528445 \\ \hline
\multirow{2}{*}{type-2} & $\theta^{\circ}$ & 0.285284 & 0.380567 & 0.571420 & 1.146307 & 2.306986 & 3.483098 & 4.675807 \\ \cline{2-9} 
 & $\theta^{\prime \circ}$ & $-0.583049$ & $-0.584172$ & $-0.586426$ & $-0.593244$ & $-0.607136$ & $-0.621395$ & $-0.636048$ \\ \hline
\end{tabular}
\caption{
The calculated ($\theta$, $\theta'$) angle sets for doubly commensurate tBG/BN
type-1 and type-2 systems where the moire patterns are aligned modulo $60^{\circ}$ angles 
when the moire patterns can be related through 
$L^M_{\rm tGBN} = N \times  L^M_{\rm tBG}$.}
\label{energy_multiple_period}
\end{table*}

\begin{figure*}
\begin{center}
\includegraphics[width=0.95\textwidth]{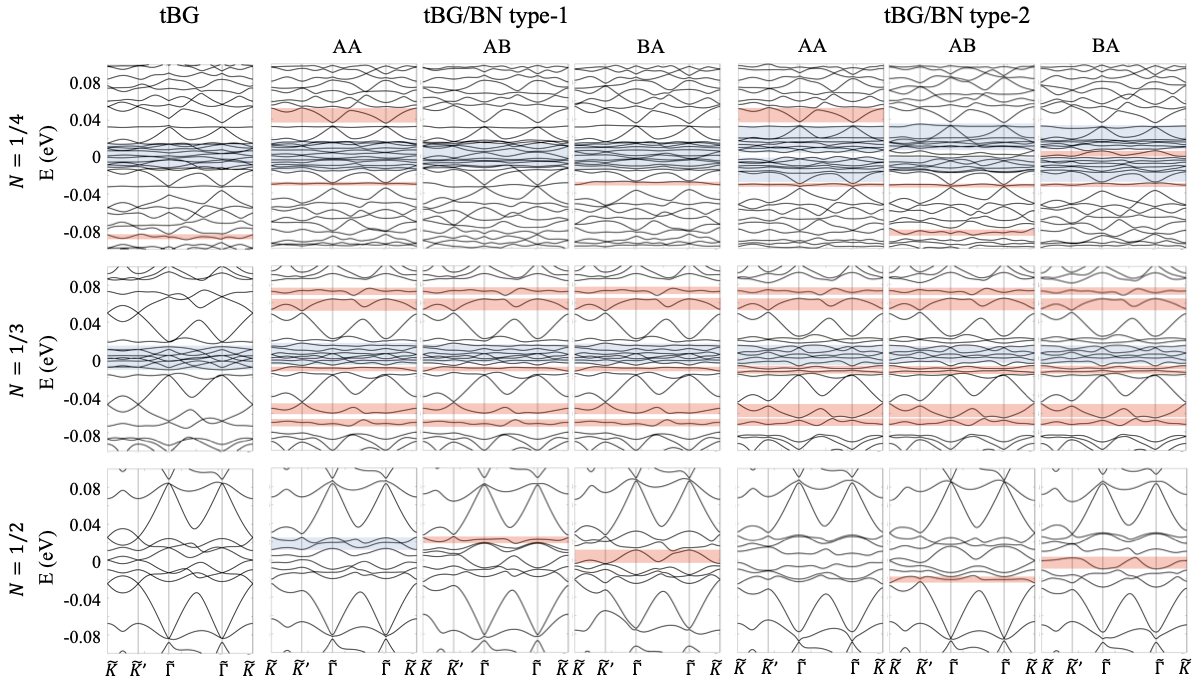}
\end{center}
\caption{
Band structures for the twist angle sets of $N$-multiple moire lengths for $N = 1/4, 1/3, 1/2$
for twist angles listed in Table~\ref{energy_multiple_period}
for single moire tBG and double moire tBG/BN  type-1 and type-2  systems. 
The AA, AB, and BA labels indicate the starting stacking of the top G layer prior to rotation. 
Isolated folded band bundles with narrow width $\lesssim 30$~meV are indicated by the blue-shaded regions, while the isolated single flat bands of width $\lesssim 20$~meV are indicated by red-shaded regions. 
	}\label{fig7_multiple}
\end{figure*}

%------------------------------------------------------------%
\section{Commensurate double moire angle sets}
So far we have considered doubly commensurate systems with equal moire pattern lengths
and here we extend our band structure analysis to commensurate double moire systems 
with unequal individual moire lengths. 
An example of a doubly commensurate supermoire built from unequal length moire patterns was realized
in twisted double bilayer graphene on hexagonal boron nitride arranged in a way that
it gives rise to a Kekule-type supermoire lattice whose moire bands are flattened
and isolated~\cite{kekuletdbg}. 
We expect that a similar bandwidth narrowing and band isolation behavior due to zone folding and formation of avoided gaps
could take place in tBG/BN systems. 
Here we explore the band structures of supermoire systems formed by combining 
unequal individual moire lengths that are multiples of each other and have modulo 60$^{\circ}$ alignment. 
The Hamiltonian matrix elements of the commensurate double moire pattern
can be built from the matrix elements of each moire pattern labeled by the 
respective moire reciprocal lattice vectors whose magnitude and relative orientations 
will change depending on the twist angle sets considered. 

Our analysis focuses on the $L^M_{\rm tBG} > L^M_{\rm tGBN}$ cases when the moire lengths of tBG are 
greater than those of tBG for $N = 1/2, 1/3,$ and $1/4$ ratios considering that
the moire length $L^M_{\rm tBG} \simeq a_{\rm G}/ \theta$ for tBG and 
$L^M_\textrm{\rm tGBN}\simeq a_{\rm G}/ (\theta'^2 + \epsilon^2)^{1/2}$ for tGBN as a function of twist angles
$\theta$, $\theta'$ are given in Eq.~(\ref{moirelen1}) and (\ref{moirelen2}). 
These commensurate supermoire patterns can be achieved in the limit of small tBG twist angles 
$\theta \lesssim 1^{\circ}$ and gives rise to zone folding as the moire Brillouin zone reduces in size. 
The set of twist angles that satisfy the commensuration conditions 
are listed in Table~\ref{energy_multiple_period} for type-1 and type-2 systems
when $L^M_{\rm tGBN}$ = $N$ $\times  L^M_{\rm tBG}$ for some integer and rational values of $N$.

The resulting band structures are presented in Fig.~\ref{fig7_multiple} for the three different starting stackings, 
AA, AB, and BA where we can observe bundles of narrow bands that we highlight with blue-shades
that are separated from each other forming bands of moire bands.
Likewise we have also highlighted with red shades the isolated nearly flat moire bands
that are separated from each other through avoided gaps.
In particular we find that the band bundles for $N = 1/4$, $1/3$ commensurate ratios 
can have bandwidths of the order of $\sim 30$~meV.

These folded nearly flat bands are weakly dispersive in the reduced moire 
Brillouin zones in the limit of small tBG twist angles.
We expect that the moire pattern features that appear at low energies 
due to the BN substrate can introduce gaps
that separate the folded band bundles that enhance their isolation and reduce their screening. 
The notable changes in the band structures that we can observe for different relative sliding configurations
for the doubly commensurate geometries indicate that small deviations of the twist angles can in principle
lead to significant changes to the electronic structures, 
while we might also expect formation of gaps between the folded moire 
band bundles due to the BN substrate. 
As we approach the regime of marginally twisted graphene bilayers we expect that the 
commensuration strain effects will become more important and should be 
considered in a more complete theory.

%------------------------------------------------------------%
\section{conclusion}\label{conclusion}

We have investigated the electronic structure of commensurate double moire patterns of twisted 
bilayer graphene (tBG) on hexagonal boron nitride (BN) in an effort to understand the effects introduced by 
a BN substrate when it is brought to near alignment with the graphene layers, 
and identify the conditions that are favorable for the appearance of gapped moire bands with 
finite valley Chern numbers that ultimately lead to the anomalous Hall effects observed in experiments.

The band structure of tBG has been calculated based on a continuum model that effectively 
accounts for vertical interlayer relaxations by using unequal interlayer tunneling parameters. 
The effects of a second moire pattern produced by the BN substrate have been modeled assuming a 
moire pattern that is perfectly commensurate with tBG
which conveniently allows us to study the physics of the system based on a single moire 
Brillouin zone continuum Hamiltonian model. 
For these doubly commensurate geometries we have highlighted the importance of aligning properly 
the moire reciprocal lattice upon rotation to properly construct the continuum moire Hamiltonian,
and have illustrated the role of the BN moire patterns and interlayer potential differences in opening the 
primary band gaps by obtaining the analytical form of the gaps near $\tilde{K}'$ for a truncated model. 

When the periods of each moire patterns are equal, we have identified two types of 
commensurate double moire patterns depending on the relative
twist angles of the tBG layers with respect to the BN substrate that we classified as type-1 and type-2
and whose band structures have revealed markedly different features in their electron-hole asymmetry,
the formation of band gaps, and dependence on local interlayer sliding arrangement.
In type-1 double moire systems we find narrower overall conduction bands than valence bands 
with the most prominent asymmetry appearing in the vicinity of the BA-tBG/BN stacking configuration 
where the conduction band has the smallest bandwidth $\sim 5$~meV suggesting that conduction bands
are more prone for broken symmetry phases than the valence bands, 
whereas the electron-hole symmetry is partially restored in type-2 solutions that slightly favor on average the narrowing of the valence bands suggesting that ordered phases will likely appear for both conduction and valence bands. 
The impact of the BN substrate on the electronic structure should be observable not only through transport measurement 
but also through the optical absorption probes as we illustrated for the specific BA-tBG/BN example versus tBG without a substrate. 

We have extended our electronic structure study to other commensurate double moire systems 
with unequal moire pattern lengths $ L^M_{\rm tBG}  >  L^M_{\rm tGBN} $ focusing on  
the smaller bilayer graphene twist angles and verified that they 
often lead to separated folded moire bands bundles with reduced widths $\sim 30$~meV.
Hence, our results suggest that tBG twist angles in the range of $0.3^{\circ} \sim 0.5^{\circ}$ clearly below $\sim 1^{\circ}$
can also be candidate systems for exhibiting correlated phases. 

The analysis presented in this work for commensurate double moire patterns
for a variety of different sliding geometries shows how the BN substrate impacts the electronic 
structre of tBG, specifically leading to band gap openings that isolate the moire bands. 
Additionally, these band structures provide a useful starting point for understanding the 
electronic structure of incommensurate double moire tBG/BN systems that can be
pictured as a coherent combination of the commensurate states with different stacking.
A more detailed study of the incommensurate double moire patterns 
as they depart from perfect commensuration as well as the impact 
of lattice relaxations in the electronic structure is desirable to understand in greater detail
the precursor states leading to the experimentally observed anomalous Hall orbital 
magnetization in tBG/BN.

\section{acknowledgments}
We gratefully acknowledge discussions with J.~H. Sun for the derivation of the analytic band gaps. 
This work was supported by Samsung Science and Technology Foundation under project 
no.~SSTF-BA1802-06 for J.~S., 
the Korean National Research Foundation (NRF) grant NRF-2020R1A5A1016518 for Y.~P., 
the Basic Science Research Program of the NRF-2018R1A6A1A06024977 for B.~L.~C., 
and by grant NRF- 2020R1A2C3009142 for J.~J. 
We acknowledge computational support from KISTI through grant KSC-2020-CRE-0072.

\renewcommand{\thetable}{A\arabic{table}}
\renewcommand{\thefigure}{A\arabic{figure}}
\renewcommand{\theequation}{A\arabic{equation}}
\setcounter{table}{0}
\setcounter{figure}{0}
\setcounter{equation}{0}

\begin{figure*}
\begin{center}
\includegraphics[width=1.0\textwidth]{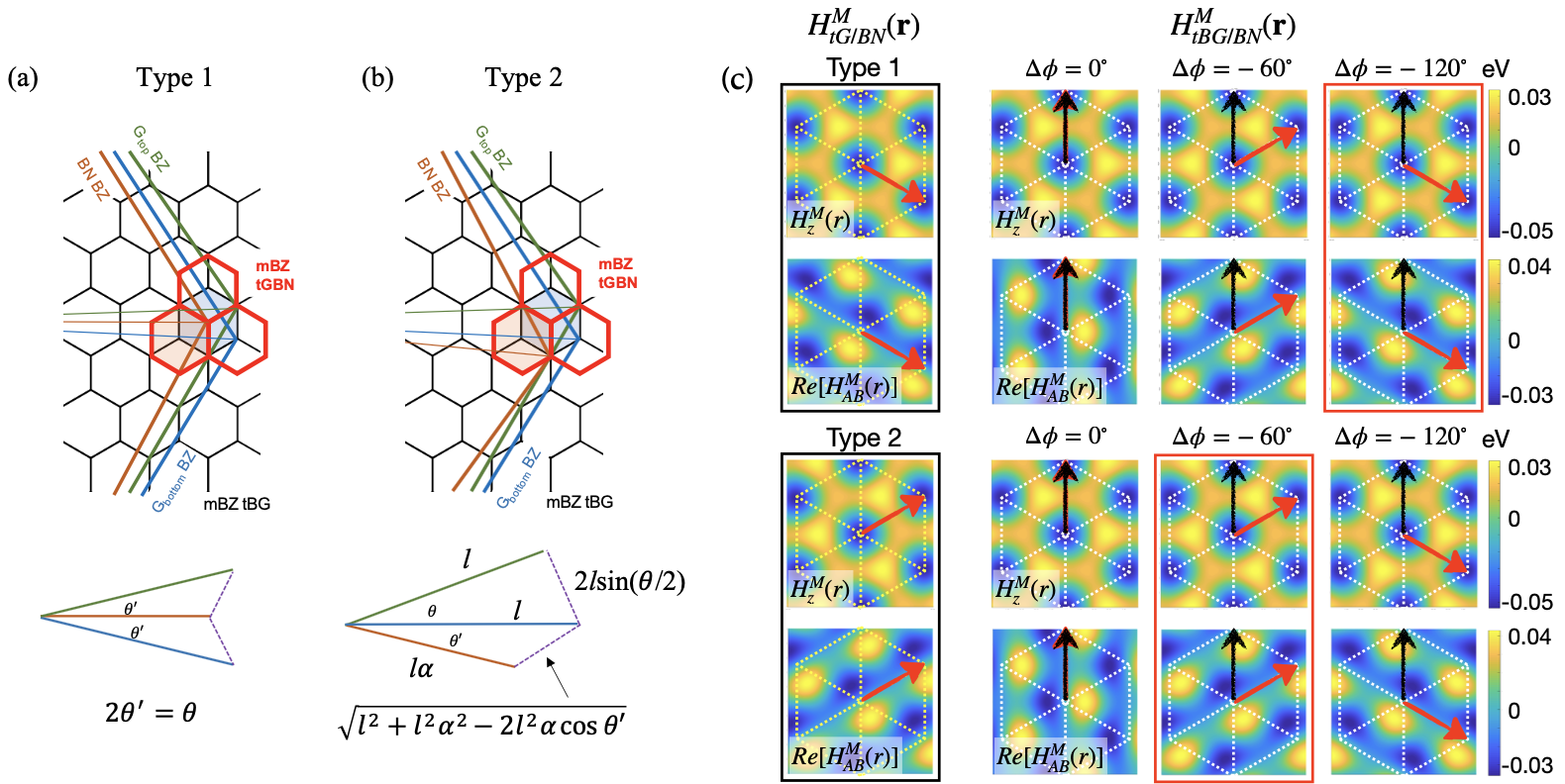}
\end{center}
\caption{
	(Color online) Geometric solution of the two sets of commensurate angles for tBG/BN in momentum-space for (a) type-1 ($\theta>0$, $\theta' > 0$) and (b) type-2 ($\theta>0$, $\theta' < 0$). The two dashed (purple) lines on the lower row for both type-1 and type-2 have the equal length. (c) The leftmost coumn represents real-space representation of mass term $H_z^{M}(\bm{r})$ and the real part of the in-plane gauge field $\textrm{Re}[H^M_{AB}(\bm{r})]$ in tGBN for type-1 (upper two rows) and type-2 (lower two rows). The red arrow indicates the lattice vector of tGBN. The rest three columns represent the mass term and the real part of the in-plane gauge term in tBG/BN for type-1 (upper two rows) and type-2 (lower two rows) for three different $\Delta \phi$ values $0^\circ$, $-60^\circ$, and $-120^\circ$. The black arrows correspond to the lattice vector of tBG and the red arrows represent that of tBG/BN. Note that the moire patterns generated from tBG/BN coincide with those from tGBN (on the leftmost column) when $\Delta \phi$ (the angle difference between red and black arrows) is $-120^\circ$ for type-1 and $\Delta \phi = -60^\circ$ for type-2. 
	}\label{Afig1}
\end{figure*}

\section*{Appendix A: Geometrical derivation of commensurate twist angles and moire pattern rotations}\label{AppendixA}
There are two classes of solutions for moire lengh $L^M_{\rm tBG} = L^M_{\rm tGBN}$ commensurate tBG/BN as shown in Fig.~\ref{fig1}. 
Here we present a geometrical proof for both types of solution in momentum-space and a more detailed analysis about how type-1 and type-2 moire potentials in real-space 
lead to relative moire pattern rotation angles of $\Delta \phi = -120^\circ$ and $\Delta \phi = -60^\circ$ respectively.
 
We assume that the bottom G and top G layers are rotated by $-\theta/2$ and $\theta/2$, respectively. 
In Fig.~\ref{Afig1} we plot the BZ of the bottom G layer with a thick blue  solid line and the BZ of the top G layer with a thick green  solid line, generating the mBZ of tBG which plotted by a black thin solid line. The thick brown solid line represents BZ of hBN. For tBG/BN to be commensurate, mBZ of tBG and of tGBN should have the same size and share their two corners. The geometrical understanding for type-1 is straightforward since the simple relation $\theta = 2\theta'$ holds as shown in Fig.~\ref{Afig1}(a). For type-2, on the other hand, if we assume the length from the center to the corner of BZ of hBN layer is $l\alpha$, then the length from the center to the corner of BZ of G layer is $l$, and the two lengths indicated by the dashed purple line in Fig. \ref{Afig1} must be the same to share the two corners of the mBZ of tBG and tGBN. Then we get an equation which is equivalent to $L^M_\textrm{tBG} = L^M_\textrm{tGBN}$ condition in 
Eq.~(\ref{relation_th_thp}). The shaded areas in red and blue in the upper row in Fig.~\ref{Afig1}(a) and (b) 
correspond to the mBZ depicted in Fig. \ref{fig1}(e) and (j), respectively.

We now move on to discuss the rotation of the moire patterns. 
The moire potential in Eq.~(\ref{moireeq}) can be rewritten for the bottom G layer with a 2$\times$2 matrix as follows

\begin{equation}
\begin{array}{ll}
\vspace{0.6cm}
H^M(\bm{r})& = \sum_{m=1}^6 W_{m} e^{i\bm{\tilde{G'}}_m \cdot \bm{r}} \\
\vspace{0.6cm}
&= V_{\rm AA}(\bm{r}) \big( \frac{\sigma_0 + \sigma_z}{2}\big) + V_{\rm BB}(\bm{r})\big( \frac{\sigma_0 - \sigma_z}{2}\big) \\
%&+ H^M_{AB}(\bm{r})
&+ \bm{A}(\bm{r}) \cdot \bm{\sigma}
\end{array}
\end{equation}
where $\bm{\tilde{G}}'_m$ are the moire reciprocal vectors of tGBN
\begin{equation}
V_{\rm AA}(\bm{r}) = 2 C_{\rm AA} \textrm{Re} [e^{i\phi_{\rm AA}}f(\bm{r})], 
\end{equation}
\begin{equation}
V_{\rm BB}(\bm{r}) = 2 C_{\rm BB} \textrm{Re} [e^{i\phi_{rm BB}}f(\bm{r})],
\end{equation}
and 
\begin{equation}
\bm{A}(\bm{r}) = 2C_{\rm AB} (\hat{z} \times \bm{\nabla} ) \textrm{Re}[e^{i \phi_{AB} }f(\bm{r})]/|\bm{\tilde{G}}|.
\end{equation}
Here, $f(\bm{r}) = \sum_m (1+(-1)^m)e^{i\tilde{\bm{G}'}_m \cdot \bm{r}}/2$ and $|\bm{\tilde{G}}|$ is the magnitude of the 
moire reciprocal lattice vector $\bm{\tilde{G}}_m$ of tBG. 
In the pseudospin basis 
$H^M_0(\bm{r}) = (V_{\rm AA}(\bm{r}) + V_{\rm BB}(\bm{r}))/2$ represents the periodic moire potential, 
and $H^M_z(\bm{r}) = (V_{\rm AA}(\bm{r}) - V_{\rm BB}(\bm{r}))/2$ is the local mass term that can 
gives a gap at the neutrality point. 
The $\bm{A}(\bm{r})$ is a non-Abelian SU(2) gauge potential in the in-plane direction due to the 
unequal in the hopping probability between different sublattices of a honeycomb lattice~\cite{VOZMEDIANO2010, Jung2017a}. We can rewrite the $\bm{A}(\bm{r}) \cdot \bm{\sigma}$ term as $H^M_{\rm AB}(\bm{r}) \sigma_+ + H^M_{\rm BA}(\bm{r}) \sigma_-$ where $\sigma_\pm$ is $\big(\sigma_x \pm i\sigma_y\big)/2$, and then
from the real space maps of the rotated moire potentials $H^M_z(\bm{r})$ and $H^M_{\rm AB}(\bm{r})$
we can verify the correct alignment required for the moire reciprocal lattice vectors of both interfaces 
to assign the moire pattern coefficients in the Hamiltonian matrix. 
%for a given type of the commensurate angle set, and expect that the resulting tGBN components are exactly the same as tBG/BN in real-space.

The leftmost column of Fig.~\ref{Afig1} (c) shows the mass term $H^M_z(\bm{r})$ in real-space and the real part of gauge field $\textrm{Re}[H^M_{\rm AB}(\bm{r})]$ of tGBN and the red arrows represent the one of moire lattice vectors resulted from the moire patterns. On the other hand, the right three columns gives $H^M_z(\bm{r})$ and $\textrm{Re}[H^M_{\rm AB}(\bm{r})]$ of tBG/BN for different $\Delta \phi$. The black arrows are the moire lattice vector directions of tBG and the red arrows are those of tBG/BN. Note in Fig. \ref{Afig1} (c) that the two components of the moire potential, $H^M_z(\bm{r})$ and $H^M_{\rm AB}(\bm{r})$ in real-space for tGBN are in agreement with those of tBG/BN when $\Delta \phi = -120^\circ (-60^\circ)$ for the case of type-1 (type-2) as also illustrated in more detail in Fig. \ref{fig1} (e), (j).

\section*{Appendix B: Band gap analysis at $\tilde{K}'$}\label{AppendixB}
Here derive the analytic expressions for the energy gap at a specific high-symmetry point $\tilde{K}'$ that 
often is a useful estimate of the actual band gap at charge neutrality. 
We first consider truncating the range of $\tilde{\bm{G}}$-vectors for an eight-bands model of tBG~\cite{bistritzer11} 
possessing three Dirac Hamiltonians for the bottom layer and one Dirac Hamiltonian for the top layer. 
We use the fact that among the four two-component pseudospin eigenstates in the basis 
$ \Psi = (\psi_1, \psi_2, \psi_3, \psi_4 )^T$, \cite{Javvaji2020} 
the two spinors $\psi_2$ and $\psi_3$ can be written at $\tilde{K}'$ in terms of 
$\psi_4$ by a unitary transformation as follows, for example, for AA-tBG/BN
\begin{equation}
\psi_2= 
\begin{pmatrix}
1&0\\
0&e^{i\varphi}\\
\end{pmatrix} \psi_4, ~~~\psi_3= 
\begin{pmatrix}
1&0\\
0&e^{i\varphi}\\
\end{pmatrix}^\dagger \psi_4
\end{equation}
for conduction (valence) band in type-1 (type-2), and
\begin{equation}
\psi_2= 
\begin{pmatrix}
e^{-i\varphi}&0\\
0&1\\
\end{pmatrix} \psi_4, ~~~\psi_3= 
\begin{pmatrix}
e^{-i\varphi}&0\\
0&1\\
\end{pmatrix}^\dagger \psi_4
\end{equation}
for valence (conduction) band in type-1 (type-2), we reduced our 8$\times$8 Hamiltonian into 3$\times$3 matrix to find a secular equation. In type-1 for the conduction band, 
\begin{widetext}

\begin{equation}
\begin{pmatrix}
-E_C & 3~e^{-i\varphi} \omega & 3 \omega^{\prime} \\
e^{i\varphi}~\omega  & 2 C_{\rm AA}{\rm cos}(\phi_{\rm AA}+\varphi) - E_C  & 2 C_{\rm AB}{\rm cos}(\phi_{\rm AB})~e^{-i\varphi} + Q~e^{i\pi/6} \\
\omega^{\prime}  & 2 C_{\rm AB}{\rm cos}(\phi_{\rm AB})~e^{i\varphi} + Q~e^{-i\pi/6} & 2 C_{\rm BB}{\rm cos}(\phi_{\rm BB}) - E_C \end{pmatrix} 
\begin{pmatrix}
c \\
a \\
b \end{pmatrix} 
=
\begin{pmatrix}
0 \\
0 \\
0 \end{pmatrix} 
\end{equation}

and for the valence band,

\begin{equation}
\begin{pmatrix}
-E_V & 3 \omega^{\prime} & 3~e^{i\varphi} \omega \\
\omega^{\prime}  & 2 C_{\rm AA}{\rm cos}(\phi_{\rm AA}) - E_V  & 2 C_{\rm AB}{\rm cos}(\phi_{\rm AB}-\varphi/2)~e^{i\varphi/2} + Q~e^{i \pi/6} \\
e^{-i\varphi}~\omega  & 2 C_{\rm AB}{\rm cos}(\phi_{\rm AB}-\varphi/2)~e^{-i\varphi/2} + Q~e^{-i\pi/6} & 2 C_{\rm BB}{\rm cos}(\phi_{\rm BB}-\varphi) - E_V \end{pmatrix} 
\begin{pmatrix}
c \\
a \\
b \end{pmatrix} 
=
\begin{pmatrix}
0 \\
0 \\
0 \end{pmatrix} 
\end{equation}

where the eigenstates of the conduction and valence band for eight-band model are given as $\psi_{C,1} = \begin{pmatrix}c &0 & a & be^{i\varphi} &a & be^{-i\varphi} &a & b\end{pmatrix}^T$ and $\psi_{V,1} = \begin{pmatrix}0 &c & ae^{-i\varphi} & b &ae^{i\varphi} & b &a & b\end{pmatrix}^T$. Here, $Q = \hbar \upsilon_F \theta k_D$.
\end{widetext}

We get characteristic equations from the above $3 \times 3$ matrices. For example, 
\begin{equation} 
E_{C}^3 - 2A_1 E_{C}^2 - A_2E_{C} + 6A_3 = 0,
\label{ceq1}
\end{equation} and 
\begin{equation}
E_{V}^3 - 2B_1E_{V}^2 - B_2E_{V} + 6B_3 = 0,
\label{ceq2}
\end{equation}
where
\begin{equation}
A_1(B_1) = C_{\rm AA} \cos \Theta^{A(B)}_{\textrm{I}, t} + C_{\rm BB} \cos \Theta^{A(B)}_{\textrm{II}, t}, 
\label{constsA1B1}
\end{equation}
and
\begin{equation}
\begin{array}{ll}
A_2 (B_2) & = 3(\omega'^2 + \omega^2) + Q^2 + 4C_{\rm AB}^2 \cos^2 \Xi_{\textrm{I}, t}^{A(B)} \\
& - 4C_{AA}C_{\rm BB} \cos \Xi_{\textrm{II}, t}^{A(B)} \cos \Theta_{\textrm{II}, t}^{A(B)} \\
& +2\sqrt{3} Q C_{\rm AB} \cos \Xi_{\textrm{I}, t}^{A(B)}
\end{array}
\label{constsA2B2}
\end{equation}
for $t = 1$ (type-1) and $t = 2$ (type-2).  For type-1,  $\Theta$ and $\Xi$ in the order of AA-, AB-, BA-tBG/BN are given as follows.

%-----------------------------------------------------%
\begin{equation}
\Theta_{\textrm{I},1}^{A} = \left\{ \begin{array}{ll}
 \phi_{\rm AA} \\
 \phi_{\rm AA} \\
 \phi_{\rm AA}, 
  \end{array} \right. \hspace{0.8cm}
  \Theta_{\textrm{I},1}^{B} = \left\{ \begin{array}{ll}
 \phi_{\rm AA} - \varphi \\
 \phi_{\rm AA} \\
 \phi_{\rm AA} -  \varphi,
  \end{array} \right. 
\end{equation}

%-----------------------------------------------------%

\begin{equation}
  \Theta_{\textrm{II},1}^{A} = \left\{ \begin{array}{ll}
 \phi_{\rm BB} + \varphi \\
 \phi_{\rm BB} - \varphi \\
 \phi_{\rm BB} - \varphi 
  \end{array} \right.\hspace{0.5cm}
  \Theta_{\textrm{II},1}^{B} = \left\{ \begin{array}{ll}
 \phi_{\rm BB} \\
 \phi_{\rm BB} + \varphi \\
 \phi_{\rm BB} 
  \end{array} \right.
\end{equation}

%-----------------------------------------------------%
\begin{equation}
\Xi_{\textrm{I},1}^{A} = \left\{ \begin{array}{ll}
 \phi_{\rm AB} \\
 \phi_{\rm AB} - \varphi \\
 \phi_{\rm AB} - \varphi, 
  \end{array} \right.  \hspace{0.8cm}
  \Xi_{\textrm{I},1}^{B} = \left\{ \begin{array}{ll}
 \phi_{\rm AB} + \varphi \\
 \phi_{\rm AB} \\
 \phi_{\rm AB} + \varphi, 
  \end{array} \right. 
  \end{equation}
  \begin{equation}
  \Xi_{\textrm{II},1}^{A} = \left\{ \begin{array}{ll}
 \phi_{\rm AA} \\
 \phi_{\rm AA} + \varphi \\
 \phi_{\rm AA} + \varphi, 
  \end{array} \right. \hspace{0.8cm}
    \Xi_{\textrm{II},1}^{B} = \left\{ \begin{array}{ll}
 \phi_{\rm AA} - \varphi\\
 \phi_{\rm AA} \\
 \phi_{\rm AA} - \varphi, 
  \end{array} \right.
\end{equation}

\begin{comment}
\begin{equation}
\Xi_{\textrm{III},1}^{A} = \left\{ \begin{array}{ll}
 \phi_{\rm BB} + \varphi \\
 \phi_{\rm BB} - \varphi \\
 \phi_{\rm BB} - \varphi, 
  \end{array} \right. \hspace{0.8cm}
  \Xi_{\textrm{III},1}^{B} = \left\{ \begin{array}{ll}
 \phi_{\rm BB} \\
 \phi_{\rm BB} + \varphi \\
 \phi_{\rm BB}, 
  \end{array} \right.
\end{equation}
\end{comment}

%-----------------------------------------------------%
Also, $A_3$ and $B_3$ are given as follows, for type-1 AA-tBG/BN,
\begin{equation}
\begin{array}{ll}
A_3 & = \omega'^2 C_{\rm BB} \cos (\phi_{\rm BB} + \varphi) + \omega^2 C_{\rm AA} \cos (\phi_{\rm AA}) \\
&+ \omega' \omega C_{\rm AB} \cos (\phi_{\rm AB}), 
\end{array}
\end{equation}
\begin{equation}
\begin{array}{ll}
B_3 &= \omega'^2 C_{\rm AA} \cos (\phi_{\rm AA} - \varphi) + \omega^2 C_{\rm BB} \cos (\phi_{\rm BB}) \\
&+ \omega' \omega C_{\rm AB} \cos( \phi_{\rm AB}+\varphi), 
\end{array}
\end{equation}
for type-1 AB-tBG/BN,
\begin{equation}
\begin{array}{ll}
A_3&=\omega^2C_{\rm AA}{\rm cos}(\phi_{\rm AA}+\varphi) + \omega'^2C_{\rm BB}{\rm cos}(\phi_{\rm BB}-\varphi)\\
&+\omega'\omega C_{\rm AB} {\rm cos}(\phi_{\rm AB}-\varphi),
\end{array}
\end{equation} 

and
\begin{equation}
\begin{array}{ll}
B_3&=\omega'^2C_{\rm AA}{\rm cos}(\phi_{\rm AA}) +\omega^2C_{\rm BB}{\rm cos}(\phi_{\rm BB}+\varphi) \\
&+ \omega'\omega C_{\rm AB}{\rm cos}(\phi_{\rm AB}),
\end{array}
\end{equation} 
and for type-1 BA-tBG/BN,
\begin{equation}
\begin{array}{ll}
A_3&=\omega'^2 C_{\rm AA}{\rm cos}(\phi_{\rm AA}+\varphi) + \omega^2 C_{\rm BB}{\rm cos}(\phi_{\rm BB}-\varphi)\\
&+\omega'\omega C_{\rm AB} {\rm cos}(\phi_{\rm AB}-\varphi),
\end{array}
\end{equation} 

and 
\begin{equation}
\begin{array}{ll}
B_3&=\omega^2 C_{\rm AA}{\rm cos}(\phi_{\rm AA}-\varphi) + \omega'^2 C_{\rm BB}{\rm cos}(\phi_{\rm BB}) \\
&+ \omega'\omega C_{\rm AB}{\rm cos}(\phi_{\rm AB}+\varphi). 
\end{array}
\end{equation}

Since $A_3$ and $B_3$ are negligibly small, we can get three roots for each, $E_{C} = 0$, $E_{C} = A_1 \pm \sqrt{A_1^2 + A_2}$, and $E_{V} = 0$, $E_{V} = B_1 \pm \sqrt{B_1^2 + B_2}$, presuming $A_3= B_3 = 0$. By retrieving $A_3$, $B_3$ and plugging the lowest energies $E_{C} = 0$, $E_{V} = 0$, whose eigenvectors remain the same, into the original characteristic equations, we get a much more simplified expressions for the energy gap as written in Eq. (\ref{analyticGap}). In the same manner, for other stackings of tBG the characteristic equations have the same form but the coefficients are different since the unitary matrix $U$ connecting the spinor bases are different, resulting in different characteristic equations. 

For AA-tBG/BN of type-2, the $3\times3$ matrix for a characteristic equation is given as
\begin{widetext}
\begin{equation}
\begin{pmatrix}
-E_C & 3 e^{-i\varphi}~\omega &3 \omega^{\prime} \\
e^{i\varphi}~\omega  & 2 C_{\rm AA}{\rm cos}(\phi_{\rm AA}+\varphi) - E_C  & 2 C_{\rm AB}{\rm cos}(\phi_{\rm AB})e^{-i\varphi} + Q~e^{i\pi/6} \\
\omega^{\prime}  & 2 C_{\rm AB}{\rm cos}(\phi_{\rm AB})~e^{i\varphi} + Q~e^{-i\pi/6} & 2 C_{\rm BB}{\rm cos}(\phi_{\rm BB}) - E_C \end{pmatrix} 
\begin{pmatrix}
c \\
a \\
b \end{pmatrix} 
=
\begin{pmatrix}
0 \\
0 \\
0 \end{pmatrix} 
\end{equation}

\begin{equation}
\begin{pmatrix}
-E_V & 3 \omega^{\prime} & 3 e^{i\varphi}~\omega \\
\omega^{\prime}  & 2 C_{\rm AA}{\rm cos}(\phi_{\rm AA}) - E_V  & 2 C_{\rm AB}{\rm cos}(\phi_{\rm AB}-\varphi/2)~e^{i\varphi/2} + Q~e^{i \pi/6} \\
e^{-i\varphi}~\omega  & 2 C_{\rm AB}{\rm cos}(\phi_{\rm AB}-\varphi/2)~e^{-i\varphi/2} + Q~e^{-i\pi/6} & 2 C_{\rm BB}{\rm cos}(\phi_{\rm BB}-\varphi) - E_V \end{pmatrix} 
\begin{pmatrix}
c \\
a \\
b \end{pmatrix} 
=
\begin{pmatrix}
0 \\
0 \\
0 \end{pmatrix} 
\end{equation}
\end{widetext} 
where the eigenstates of the conduction and valence band for eight-bands model are given as 
$\psi_{C,2} = \begin{pmatrix}0 &c & ae^{-i\varphi} & b &ae^{i\varphi} & b &a & b\end{pmatrix}^T$ and $\psi_{V,2} = \begin{pmatrix}c &0 & a & b e^{i\varphi} &a & b e^{-i\varphi} &a & b\end{pmatrix}^T$.  
This $3\times3$ matrix gives us two equations, Eq. (\ref{ceq1}) and Eq. (\ref{ceq2}), as in type-1. 
The constants $A_1$, $B_1$, $A_2$, $B_2$ also follow the same equations Eq.~(\ref{constsA1B1}) and 
Eq.~(\ref{constsA2B2}) with different $\Theta$ and $\Xi$ as follows.
%-----------------------------------------------------%
\begin{equation}
\Theta_{\textrm{I},2}^{A} = \left\{ \begin{array}{ll}
 \phi_{\rm AA} +\varphi\\
 \phi_{\rm AA} -\varphi\\
 \phi_{\rm AA}+\varphi, 
  \end{array} \right. \hspace{0.8cm}
  \Theta_{\textrm{I},2}^{B} = \left\{ \begin{array}{ll}
 \phi_{\rm AA} \\
 \phi_{\rm AA} \\
 \phi_{\rm AA} -  \varphi,
  \end{array} \right. 
\end{equation}

%-----------------------------------------------------%

\begin{equation}
  \Theta_{\textrm{II},2}^{A} = \left\{ \begin{array}{ll}
 \phi_{\rm BB}\\
 \phi_{\rm BB} + \varphi \\
 \phi_{\rm BB} 
  \end{array} \right.\hspace{0.5cm}
  \Theta_{\textrm{II},2}^{B} = \left\{ \begin{array}{ll}
 \phi_{\rm BB} -\varphi\\
 \phi_{\rm BB} - \varphi \\
 \phi_{\rm BB} +\varphi
  \end{array} \right.
\end{equation}

%-----------------------------------------------------%
\begin{equation}
\Xi_{\textrm{I},2}^{A} = \left\{ \begin{array}{ll}
 \phi_{\rm AB} \\
 \phi_{\rm AB} + \varphi/2 \\
 \phi_{\rm AB}, 
  \end{array} \right.  \hspace{0.6cm}
  \Xi_{\textrm{I},2}^{B} = \left\{ \begin{array}{ll}
 \phi_{\rm AB} - \varphi/2 \\
 \phi_{\rm AB} - \varphi/2\\
 \phi_{\rm AB} + \varphi/2, 
  \end{array} \right. 
  \end{equation}
  
  \begin{equation}
  \Xi_{\textrm{II},2}^{A} = \left\{ \begin{array}{ll}
 \phi_{\rm AA} +\varphi\\
 \phi_{\rm AA} - \varphi \\
 \phi_{\rm AA} + \varphi, 
  \end{array} \right. \hspace{0.8cm}
    \Xi_{\textrm{II},2}^{B} = \left\{ \begin{array}{ll}
 \phi_{\rm AA} \\
 \phi_{\rm AA} \\
 \phi_{\rm AA} - \varphi, 
  \end{array} \right.
\end{equation}

%-----------------------------------------------------%

The $A_3$ and $B_3$ parameters for type-2 systems are  given by
\begin{equation}
\begin{array}{ll}
A_3&=\omega'^2 C_{\rm AA}{\rm cos}(\phi_{\rm AA}+\varphi) + \omega^2 C_{\rm BB}{\rm cos}(\phi_{\rm BB})\\
&+\omega'\omega C_{\rm AB} {\rm cos}(\phi_{\rm AB}),
\end{array}
\end{equation} 
\begin{equation}
\begin{array}{ll}
B_3&=\omega^2 C_{\rm AA}{\rm cos}(\phi_{\rm AA}) + \omega'^2 C_{\rm BB}{\rm cos}(\phi_{\rm BB}-\varphi) \\
&- \omega'\omega C_{\rm AB}{\rm cos}(\phi_{\rm AB}-\varphi/2). 
\end{array}
\end{equation} 

for AA-tBG/BN, 
\begin{equation}
\begin{array}{ll}
A_3&=\omega^2 C_{\rm AA}{\rm cos}(\phi_{\rm AA}-\varphi) + \omega'^2 C_{\rm BB}{\rm cos}(\phi_{\rm BB}+\varphi)\\
&-\omega'\omega C_{\rm AB} {\rm cos}(\phi_{\rm AB}+\varphi/2), 
\end{array}
\end{equation} 
\begin{equation}
\begin{array}{ll}
B_3&=\omega'^2 C_{\rm AA}{\rm cos}(\phi_{\rm AA}) + \omega^2 C_{\rm BB}{\rm cos}(\phi_{\rm BB}-\varphi) \\
&- \omega'\omega C_{\rm AB}{\rm cos}(\phi_{\rm AB}-\varphi/2). 
\end{array}
\end{equation} 
for AB-tBG/BN and 
\begin{equation}
\begin{array}{ll}
A_3&=\omega^2 C_{\rm AA}{\rm cos}(\phi_{\rm AA}+\varphi) + \omega'^2 C_{\rm BB}{\rm cos}(\phi_{\rm BB})\\
&+\omega'\omega C_{\rm AB} {\rm cos}(\phi_{\rm AB}), 
\end{array}
\end{equation} 
\begin{equation}
\begin{array}{ll}
B_3&=\omega'^2 C_{\rm AA}{\rm cos}(\phi_{\rm AA}-\varphi) + \omega^2 C_{\rm BB}{\rm cos}(\phi_{\rm BB}+\varphi) \\
&- \omega'\omega C_{\rm AB}{\rm cos}(\phi_{\rm AB}+\varphi/2)
\end{array}
\end{equation} 
for BA-tBN/BN.

The analytical values of the gaps in the presence of an interlayer potential difference $\eta$ 
will be shown only 
for the case of BA-tBG/BN in type-1. The associated $3 \times 3$ matrix for the characteristic equation is 
\begin{widetext}
\begin{equation}
\begin{pmatrix}
\eta/2-E_{C} & 3e^{i \varphi}~\omega & 3 e^{-i\varphi} \omega' \\
e^{-i\varphi}\omega  & 2 C_{\rm AA}{\rm cos}(\phi_{\rm AA}+\varphi) -\eta/2- E_{C}  & 2 C_{\rm AB}{\rm cos}(\phi_{\rm AB}-\varphi) + Q~e^{i \pi/6} \\
e^{i\varphi} \omega'  & 2 C_{\rm AB}{\rm cos}(\phi_{\rm AB}-\varphi) + Q~e^{-i\pi/6} & 2 C_{\rm BB}{\rm cos}(\phi_{\rm BB}-\varphi) -\eta/2- E_{C} \end{pmatrix} 
\begin{pmatrix}
c \\
a \\
b \end{pmatrix} 
=
\begin{pmatrix}
0 \\
0 \\
0 \end{pmatrix} 
\end{equation} 
for conduction band, and
\begin{equation}
\begin{pmatrix}
\eta/2-E_{V} & 3e^{-i\varphi} \omega' & 3 \omega \\
e^{i\varphi} \omega'  & 2 C_{\rm AA}{\rm cos}(\phi_{\rm AA}-\varphi) -\eta/2- E_{V}  & 2 C_{\rm AB}{\rm cos}(\phi_{\rm AB}+\varphi) + Q~e^{i\pi/6} \\
\omega  & 2 C_{\rm AB}{\rm cos}(\phi_{\rm AB}+\varphi) + Q~e^{-i\pi/6} & 2 C_{\rm BB}{\rm cos}(\phi_{\rm BB}) -\eta/2- E_{V} \end{pmatrix} 
\begin{pmatrix}
c \\
a \\
b \end{pmatrix} 
=
\begin{pmatrix}
0 \\
0 \\
0 \end{pmatrix} 
\end{equation} 
\end{widetext} 
for the valence band, where the eigenstates of the conduction and valence bands 
in the eight-bands model are given as $\psi_{C,1} = \begin{pmatrix}0 &c & ae^{i\varphi} & be^{-i\varphi} &ae^{-i\varphi} & be^{i\varphi} &a & b\end{pmatrix}^T$ and $\psi_{V,1} = \begin{pmatrix}c &0 & ae^{-i\varphi} & b &ae^{i\varphi} & b &a & b\end{pmatrix}^T$. 
This expressions is the same as Eq.~(\ref{analyticGap}) for the gap at $\tilde{K}'$ but using with different coefficients.

\begin{figure}
\begin{center}
\includegraphics[width=0.44\textwidth]{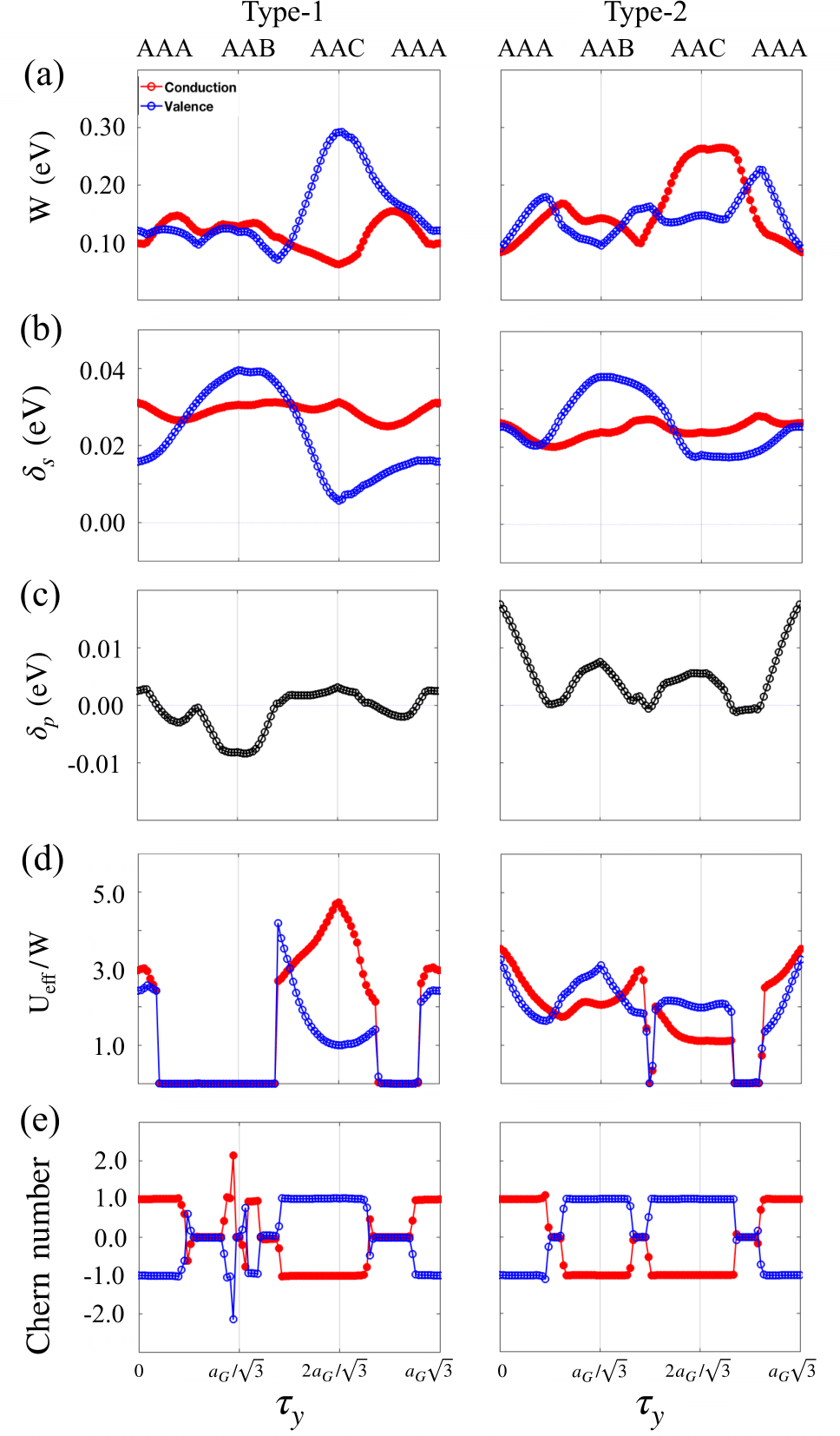}
\end{center}
\caption{
	(Color online) (a) Bandwidth W, (b) secondary gaps $\delta_s$, (c) primary gaps $\delta_p$, (d) U$_\textrm{eff}/W$ and (e) valley Chern number of the lowest conduction (red filled circle) and valence band (blue empty circle) for type-1 (left column) and type-2 (right column) as a function of sliding in $y$-direction $\tau_y$. Here, AAA stacking corresponds to AA-tBG/BN [$\tau = (0, ~0)$], AAB stacking is AB-tBG/BN [$\tau = a_G(0, ~1/\sqrt{3})$], and AAC stacking corresponds to BA-tBG/BN [$\tau = a_G(0, ~2/\sqrt{3})$]. 
	}\label{fig_bw_tau}
\end{figure}

\section*{Appendix C: supplemental electronic structure, local density of states and Berry curvatures}
\label{lineplot}

Here we supplement information on electronic structure presented in the main text
for the bandwidth $W$, the secondary band gaps $\delta_s$, primary band gaps $\delta_p$, 
the ratio of the screened Coulomb interaction strength to the bandwidth $U/W$, and valley Chern numbers
by considering the effects of twist angles under the assumptions of commensurate double moire patterns
and also provide information of the local density of states at the van Hove singularities
and the Berry curvatures of the nearly flat bands. 

We begin by presenting in Fig.~\ref{fig_bw_tau} through 1D plots various electronic structure 
information for variable $\tau_y$ and fixed $\tau_x = 0$ that had been discussed in the main text.
The topmost row (a) panel shows the fluctuations in the bandwidth for the low energy conduction and 
valence bands as a function of $\tau_y$ where we can find a relatively narrower conduction band 
versus valence band for type-1 systems and more electron-hole symmetric bands for type-2 systems. 
Subsequent rows represent (b) the secondary band gaps and (c) the primary band gaps.
Information from (a)-(c) panels are used to determine (d) the $U_{\rm eff}/W$ ratio where we can clearly 
observe its relative enhancement for conduction bands in type-1 but maintaining comparable magnitudes 
for electrons and holes for type-2 systems. 
The valley Chern numbers in (e) showing well quantized values or oscillating with non-integer values
as a function of $\tau_y$ are in keeping with the fact that the bands are isolated when we have
positive primary and secondary gaps.

\begin{figure*}
\begin{center}
\includegraphics[width=1.0\textwidth]{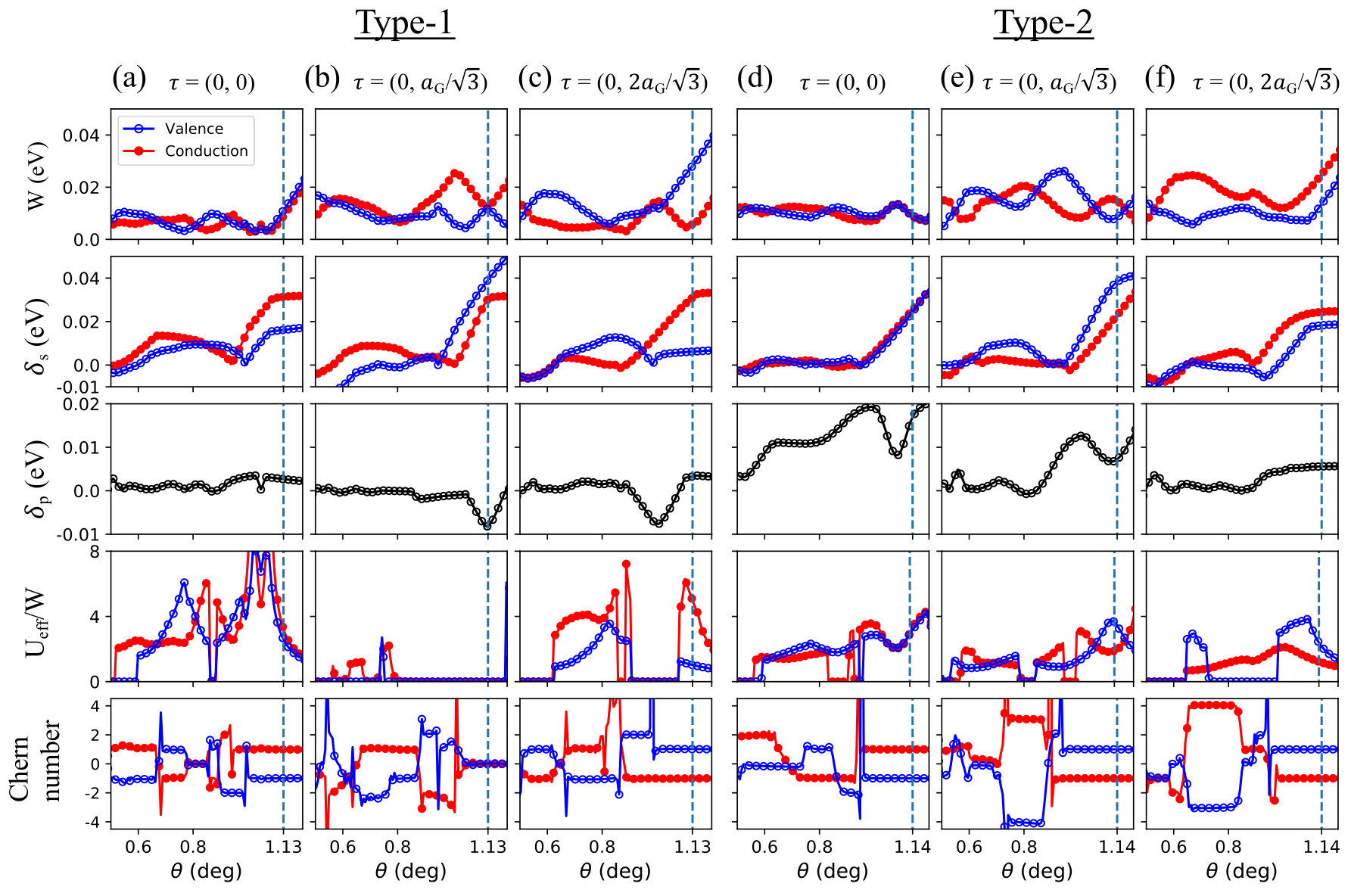}
\end{center}
\caption{
(Color online) Electronic structure of tBG/BN as a function of tBG twist angle $\theta$ for commensurate double moire patterns. 
The dashed vertical line indicates the twist angle $\theta$ corresponding to the doubly commensurate moire pattern conditions determined by the fixed lattice constants $a_{\rm G} = 2.461~\AA$ of graphene and $a_{\rm BN} = 2.504~\AA$ of hexagonal boron nitride. 
For other twist angles we assume that the twist angle $\theta'$ as well as the lattice constant of 
hBN in the tGBN moire interface changes such that the moire vectors coincide with those of tBG in both length and angle. 
From top to bottom rows we show the bandwidths, secondary gaps, primary gaps, ratio of effective Coulomb interaction to 
bandwidth (U$_\textrm{eff}/W$) of the lowest conduction (red filled circle) and valence band 
(blue empty circle) for three different stacking configurations,  AA-tBG/BN [$\tau = (0, ~0)$], 
AB-tBG/BN [$\tau = a_G(0, ~1/\sqrt{3})$], and BA-tBG/BN [$\tau = a_G(0, ~2/\sqrt{3})$] 
prior to rotation for type-1 systems in panels (a-c) and panels  (d-f) for type-2 systems.
	}
\label{fig_bw_ueff_cnum}
\end{figure*}

\begin{figure*}
\begin{center}
\includegraphics[width=1.0\textwidth]{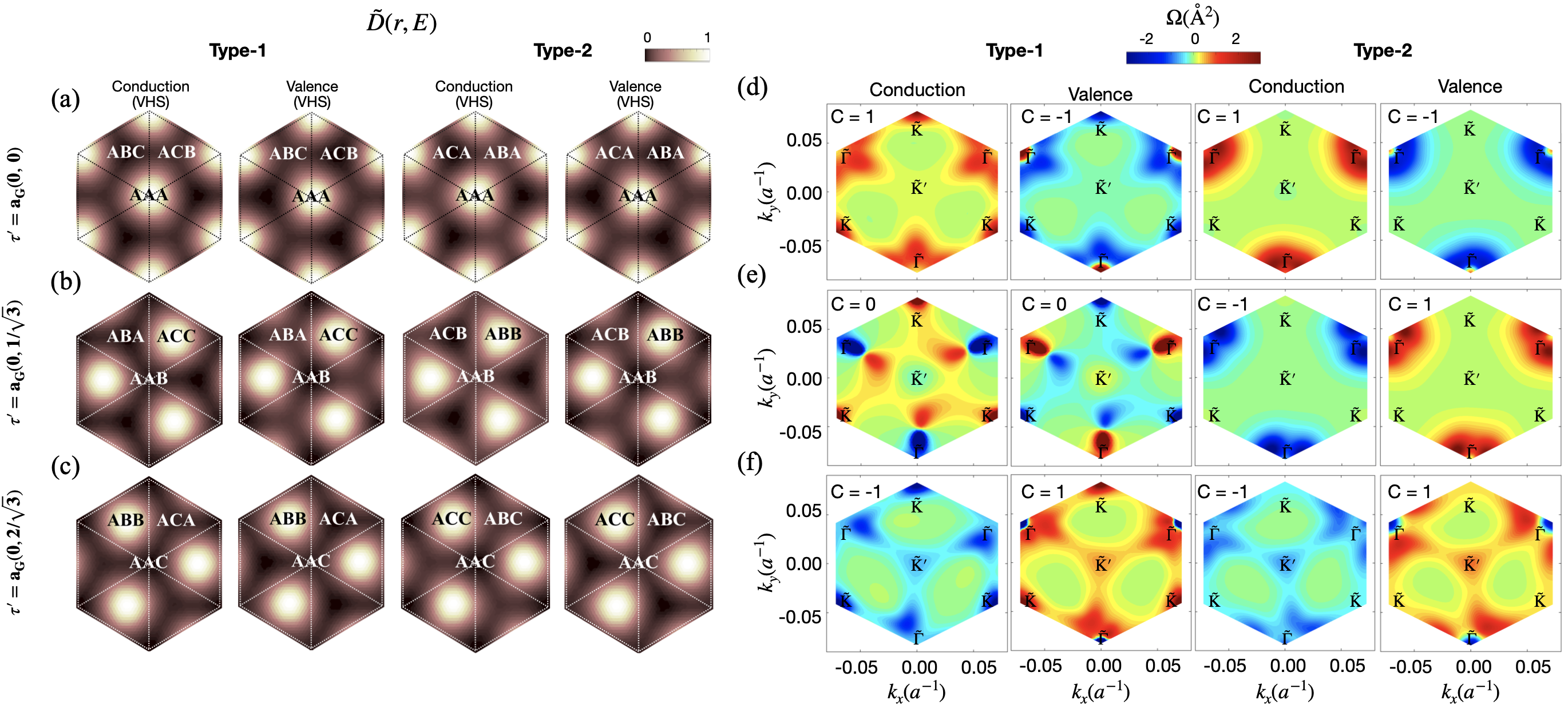}
\end{center}
\caption{
	(Color online) (Left) 
	The normalized local density of states $\tilde{D}(\bm{r}, E) = D(\bm{r}, E) / {\rm max} \vert D(\bm{r}, E) \vert$ at the 
	van Hove singularity of conduction/valence bands for the type-1 (left two columns) and type-2 (right two columns) 
	of tBG/BN.  The electrons are highly populated at the local-AA stacking in tBG for both valence and 
	conduction band of type-1 and type-2 for all three slidings as seen in (a) $\bm{\tau} =a_G(0,~0)$, 
	(b) $\bm{\tau} =a_G(0,~1/\sqrt{3})$ 	and (c) $\bm{\tau} =a_G(0,~2/\sqrt{3})$. 
	(Right) Berry curvatures of type-1 (left two columns) and type-2 (right two columns) of valence band 
	and conduction band for case of sliding the top G layer by (d) $\bm{\tau}_y = 0$, (e) $\bm{\tau}_y = a_G/\sqrt{3}$, 
	and (f) $\bm{\tau}_y = 2a_G/\sqrt{3}$.
	}
	\label{fig_ldos_tau}
\end{figure*}

In Fig.~\ref{fig_bw_ueff_cnum} we provide additional information about how the electronic structure will be modified
when we allow the twist angle $\theta$ of tBG to change from the specific values determined by the double moire 
commensuration condition. For this purpose we make the approximation that double commensuration is maintained for every value of $\theta$ which implies that the moire pattern period and angle also change continuously together with the pattern of tBG. The dashed vertical line indicates the commensuration angle $\theta$ for tBG consistent with 
the lattice constants of graphene $a_{\rm G} = 2.461~\AA$ and hexagonal boron nitride $a_{\rm BN} = 2.504~\AA$.
We may expect that the double commensuration condition can still hold approximately in the vicinity of this 
vertical dashed line thanks to lattice relaxation effects that rotate and globally relax the lattices.
The information in Fig.~\ref{fig_bw_ueff_cnum} allows to distinguish the expected electronic structures features
in the lowest energy conduction and valence bands for different initial interlayer sliding stacking geometries
AA-tBG/BN [$\tau = (0, ~0)$], AB-tBG/BN [$\tau = a_G(0, ~1/\sqrt{3})$], and BA-tBG/BN [$\tau = a_G(0, ~2/\sqrt{3})$] prior to rotation. 
While the relative sliding of the layers are shown to considerably modify the low energy electronic structure,
as we illustrate in Fig.~\ref{fig_ldos_tau} through normalized local density of states (LDOS) 
we also find that the electron localization largely concentrates at the 
AA stacking sites of tBG regardless of the arrangement of the BN substrate.  
Quantitative differences in the normalized LDOS are still expected to be observable 
near AB or BA local stacking regions when relative sliding between layers are introduced. 
Likewise the Berry curvature maps undergo changes in their hotspot distributions
depending on which stacking configuration is chosen. These differences could impact 
the Hall transport and optical measurements especially when valley contrasting
circularly polarized light is used.

% --- Esto mejor en el apendice ---

\bibliography{tbgbn}

\end{document}